\documentclass[aps,prb,reprint,superscriptaddress,amsmath,amssymb]{revtex4-1}
\usepackage{amsmath,amssymb}
\usepackage{graphicx}

\begin{document}
\title{Advances in ab-initio theory of Multiferroics\\
Materials and mechanisms: modelling and understanding}
\author{Silvia Picozzi and Alessandro Stroppa}

\affiliation{Consiglio Nazionale delle Ricerche, CNR-SPIN U.O.S. L'Aquila, Italy}

\begin{abstract}
Within the broad class of multiferroics (compounds showing a coexistence of 
 magnetism and ferroelectricity),
we focus on the subclass of ``improper electronic  ferroelectrics", {\em i.e.} correlated materials
where electronic degrees of freedom (such as spin, charge or orbital) drive  ferroelectricity.
In particular, in spin-induced ferroelectrics, 
 there is not only a {\em coexistence} of the two intriguing magnetic and dipolar orders; rather, there is such an intimate link that one drives the other, suggesting a giant magnetoelectric coupling.  
Via first-principles approaches based on density functional theory, we review the microscopic mechanisms at the basis of multiferroicity in several compounds, ranging from transition metal oxides to organic multiferroics (MFs) to organic-inorganic hybrids. 
\end{abstract}
\maketitle
\section{Introduction}

\label{intro}

Materials where cooperative phenomena - such as a switchable  
long-range dipolar or magnetic ordering or a structural deformation - spontaneously emerge below a critical temperature
are termed ``ferroic"\cite{schmidt}. Compounds, where more than one kind of ferroic order are established, are consequently denoted 
as ``multiferroics"\cite{review1,review2,review3,review4,review5}. In this Colloquium paper,  we will  concentrate on  materials showing a coexistence between (anti-)ferroelectricity and (anti-)ferromagnetism. In those systems, many degrees of freedom are simultaneously active with compe\-ting energy sca\-les, which in turn make
 the response of multiferroics to external stimuli (electric and magnetic fields, pressure, strain, doping, ...) unusually large. Indeed, multiferroics offer a wide playground for colossal cross-coupled effects to emerge.  What is generally meant  by ``cross-coupling" is a physical response {\em not} induced  by its conjugate field. Examples are   magnetoelectricity (change in magnetic properties induced by electric fields or in ferroelectric (FE) properties induced by magnetic fields), piezo\-elec\-tricity (ch\-an\-ge in structural properties induced by electric fields or  in ferroelectric properties induced by structural deformation), magnetostriction (change  in magnetic properties induced by structural deformation or in structural properties induced by magnetic fields)  
 
 Due to their multifunctional nature, multiferroics hold great potential for future technological applications (such as sensors, memories, actuators,
 switches)\cite{scott,ramesh}.
 At the sa\-me time, their complexity poses serious challenges for modelling: it calls for an accurate treatment of correlated 3d- or 4f-electrons 
and excited states, as well as for a careful description of the delicate coupling between electronic (spin, charge, orbital) degrees of freedom and structural distortions and crystal symmetries. Multiferroics therefore constitute one of the most interesting though challenging classes in modern materials 
science.
 \begin{figure}[!h]
\centerline
{\resizebox{0.62\columnwidth}{!}{%
\includegraphics{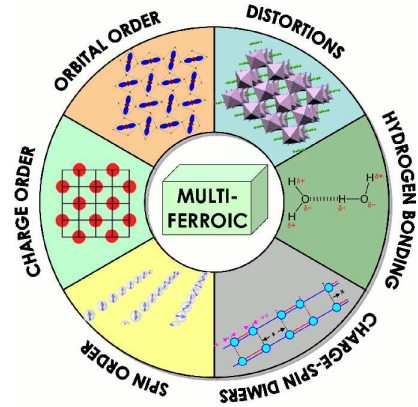}
}}
%\vspace{5cm}       % Give the correct figure height in cm
%%%%\centerline{
%           \scalebox{0.8}{
 %            \input{review_fig_mf.eps}
 %            }
 %          }

\caption{Pictorial representation of the different microscopic mechanisms that can lead to multiferroicity.}
\label{fig:cerchio}       % Give a unique label
\end{figure}

Magnetism is likely one of the earliest discovered phenomenon in condensed matter, with lodestone properties already known around 600 BC; ferroelectricity, on the other hand, is a property found rather recently, with its
earliest observation probably going back to 1921 when Valasek
 observed an  electric hysteresis in Rochelle salt\cite{valasek}.  The combination between ferroelectricity and magnetism, {\em i.e.} multiferroicity,  was
 for long time considered to be a very rare phenomenon\cite{nahill}. This common belief mostly derived  from an early 
observation\cite{cohen}: most of the ferroelectric oxides ({\em i.e.} BaTiO$_3$, PbTiO$_3$, LiNbO$_3$) have a perovskite--like structure where the  perovskite  B-site cation, which is the one that mostly off-centers and gives rise to ferroelectricity, shows an electronic $d^0$ configuration ({\em i.e.} Ti$^{4+}$, Nb$^{5+}$). 
The empty $d$-shell - required for ferroelectricity - 
 therefore seemed to preclude any coexistence 
with magnetism.  More generally and very recently, the 
the conditions of multiferroicity in $d^{n}$ perovskites are 
derived from the pseudo Jahn-Teller effect, due
to which ferroelectric displacements are triggered by 
vibronic coupling between ground and excited
electronic states of opposite parity but same spin 
multiplicity; it takes place for some specific $d^{n}$
configurations and spin states only. 
In combination with the high-spin–low-spin crossover effect this
leads to a novel phenomenon, the magnetic-ferroelectric 
(multiferroics) crossover which predicts
magnetoelectric effects with exciting functionalities 
including electric magnetization and
demagnetization.\cite{Bersuker}

``Conventional" multiferroics (whose class prototype is  
BiFeO$_3$\cite{bfo1,bfo2,bfo3}), are 
materials where, \textit{e.g.}, lone-pair electrons of the
 A-site cation in the perovskite structure 
(such as Bi $s$ with high polarizability) 
gives origin to the ferroelectric order, 
whereas the magnetic order is determined by the exchange interaction among  uncompensated spins of the B-site cation. When this conventional approach is pursued, it follows that magnetism and ferroelectricity come from two different atomic sublattices and have a chemically different origin, thereby resulting in different ordering temperatures and presumably a small magnetoelectric (ME) coupling (although this is not 
always the case\cite{bfo3}). 
These drawbacks could be overcome in materials where magnetism and ferroelectricity would share the same microscopic origin ({\em i.e.} 
the two phenomena physically originating from the same chemical species) and where the expected coupling would therefore be much stronger than in
 conventional multiferroics. Indeed, materials like those attracted an incredible attention in the last decade and will be the main topic of the present review.
 
 Being multiferroics a rather new field of materials science, in most cases the nature of the coupling between ferroelectricity and magnetism is unknown and the microscopic mechanisms at its basis have to be discovered.  Very often,  the  magnetoelectric coupling  is mediated by different degrees of freedom and interactions (orbital ordering, spin-orbit coupling, charge disproportionation, etc), therefore becoming rather complex to unveil. 
 In this context, density functional theory\cite{DFT} can be of para\-mount importance, since this ``ab-initio" approach is in principle able to describe all the many active degrees of freedom within the same level of accuracy, at variance with model-Hamiltonian approaches where a choice has to be made from the very beginning about the  relevant interactions to be taken into account. 
 
 The challenging issue of multiferroicity motivated our recent theoretical activity, which has been mainly focused on investigating microscopic mechanisms that could lead to ferroelectricity induced by breaking of inversion symmetry ({\em i.e.} the necessary symmetry condition for ferroelectricity to develop) via {\em electronic degrees of freedom}, such as charge, spin, 
orbital order\cite{review1,review4}.
In these  latter  cases,  in the framework of phase transitions, ferroelectricity is said to be ``{\em improper}", meaning that ferroelectric polarization appears  as a ``secondary" order parameter driven by a ``primary" electronic order parameter. An example will make things clearer: if one considers a (primary) antiferromagnetic (AFM) phase transition in which the AFM ordering pattern  breaks inversion symmetry, then polarization (secondary order parameter) is allowed only when the AFM order (primary order parameter) sets in. This is what happens, for example, 
in ortho-HoMnO$_3$\cite{prlslv,sergienko2007}, TbMn$_2$O$_5$\cite{laurent}, 
Ca$_3$CoMnO$_6$\cite{ca3comno6}. 
Incidentally, we here remark that conventional ferroelectric 
oxides show ``{\em proper}" ferroelectricity:
 polarization is the only order parameter in the 
phase transition and is not driven by (nor drives)
 any other electronic phase transition. 

In what will follow, we will generally consider {\em improper} ferroelectricity as {\em electronic} ferroelectricity, {\em i.e.} driven by a spin, charge or orbital phase transition. In this regard, we note that
%with the following two considerations to be taken into account: {\em i})
{\em improper} ferroelectricity can occur even when the primary order parameter is a (non-polar) 
structural distortion (as already well investigated in Refs.\cite{ghosez}), and not necessarily an electronic order parameter.

Further related to electronic ferroelectricity, we'd like to make the following comment.
 Polarization can be defined as the sum of an {\em electronic} and of an {\em ionic} contribution,  
which can be roughly understood as resulting from the polar charge rearrangement and polar distortions, respectively. Indeed, the definition of polarization is rather tricky, in particular in a periodic system. We'll here only recall some of the relevant points   and we refer the interested reader to 
Refs.\cite{compu19,compu20,compu21} where the topic is treated in a deeply detailed way.  A rough way of defining polarization is through the so called ``point-charge model" (PCM), obtained as the sum of the displacements of the atoms (with respect to a reference centrosymmetric paraelectric (PE) structure), each multiplied by the nominal valence of the corresponding ion. In this case, one adopts a purely ionic model and therefore neglects the details of the real charge distribution resulting from hybridization and covalent effects.  A rigorous way to overcome the difficulties in defining polarization is therefore provided in 
Refs.\cite{compu19,compu20,compu21} by two equivalent quantum-mechanical approaches, one based on the Berry phases and one based on Wannier function centers. According to the latter, 
the electronic charge is considered to be localized at the Wannier function centers, while the ionic charges reside  on the nuclear positions, so that
the change in electronic polarization can be obtained as a vectorial sum of the displacement of the Wannier-function centers. %with respect to the ionic positions (``anomalous"  electronic contribution) summed to the contributions due to the point-charge
%model. 
Following this approach, it is in principle possible to have an electronic contribution to polarization, due exclusively to Wannier center displacements (say, following a non-centrosymmetric AFM arrangement) on top of a centrosymmetric atomic configuration ({\em i.e.} that would otherwise lead to a vanishing contribution within simpler classical models, such as PCM). 
We however remark that
 there cannot be, strictly speaking, a purely {\em electronic} ferroelectricity, since polar atomic displacements   induced by a polar electronic order   are always present, no matter how small. Furthermore, one of the information that can actually be provided by a first-principles approach used by the present authors is indeed the quantification of electronic vs ionic contribution to polarization,  as discussed below.
 %Seen in a different way, even in the most conventional and prototypical ferroelectric oxide, BaTiO$_3$, the electronic contribution is indeed very large
%and polarization cannot be accurately estimated only by means of polar atomic displacements. 
In summary, what we mean in this Colloquium by ``electronic ferroelectricity" is the phenomenon by which  a  {\em primary electronic phase transition breaks inversion symmetry and gives, as a by-product, a ferroelectric polarization}.

A general characteristic of electronic ferroelectricity is that the expected magnitude of the polarization is likely smaller than in standard ferroelectrics; in the  case of ``improper" ferroelectricity, in fact, the polarization is driven  by a polar electronic charge rearrangement, according to which - as a by-product - the atoms are slightly displaced in a non-centrosymmetric way. On the other hand, in the latter ``proper'' case ({\em i.e.} BiFeO$_3$), the inversion symmetry is broken primarily by structural distortions, involving ionic displacements of the order of a tenth of an Angstrom (10 or 100 times larger than in electronic improper ferroelectrics). However, electronic ferroelectrics might offer a significant advantage over standard ferroelectrics, as for what concerns the switching time-scale and the so called ``fatigue". In general, the latter term means the deterioration of the hysteresis loop  and the decrease of switching charge after many polarization reversal cycles. During switching processes in proper ferroelectrics, it is the (slow and heavy) ions that displace, whereas in improper electronic ferroelectrics it is the (fast and light) electrons that move: this fundamental difference  is definitely expected to lead to a much  quicker 
switching time\cite{alexe,fiebig,dafiebig} and to a strong reduction of fatigue-related problems (although the origin of the latter is presently not well understood).
 
 As for the relevant materials, the natural class  where one expects multiferroicity to arise is represented by transition metal oxides, since they are probably the richest class in materials science where structural and electronic degrees  of freedom are all simultaneously active and interacting. However, over the years, we showed that nice effects can be found in organic materials 
as well\cite{ttfca,nature}, where various correlation phenomena usually show up and, additionally, low dimensionality can play a relevant role and can be considered as an additional degree of freedom to be tuned and exploited. Finally, very recently we discovered multiferroicity in 
organic-inorganic hybrids\cite{mof15}, such as perovskite-based metal-organic frameworks, where the unlimited variety of organic functional groups is nicely joined to the rich functionality arising from the perovskite network. 
 
 In this Colloquium paper, we put the emphasis on the rich collection of microscopic mechanisms that can lead to ``electronic ferroelectricity" , each of them manifestly at play in a different material and highlighted by means of state-of-the-art first-principles calculations in
 the different Sections. A pictorial representation of the different possible origins of multiferrocity is shown in 
 Figure\ \ref{fig:cerchio}. In particular, following a Section with computational technicalities 
(see Section\ \ref{sec:1}), we will first discuss how the spin ordering can break inversion symmetry 
(see Section\  \ref{sec:spin}), by introducing general mechanisms based on the Heisenberg symmetric exchange and Dzyaloshinskii-Moriya antisymmetric exchange. Specific examples will be discussed, based on the Heisenberg exchange involving exchange coupling between 4$f$ and 3$d$ electrons in orthoferrites (see Section\ \ref{sec:dfo}) and 
V-V exchange coupling in vanadium-based spinel (see Section\ \ref{sec:cvo}). In Sec.\ref{sec:CO} we will discuss how a specific charge-ordering can give rise to a polarization, either combined with spin-ordering or by itself, as extensively reported for magnetite in Sec.\ref{sec:magnetite}. Orbital order can also in principle break inversion symmetry and induce ferroelectricity; however, we are not presently aware of any material in which OO is the primary and unique cause of polarization (with the possible exception of double--layer manganite, although the origin of ferroelectricity is still not well understood in that compound). We  argue, anyway,  that OO can be an important ingredient which cooperates with hydrogen bonding network to induce ferroelectricity in a Cu-based metal-organic framework, as reported in Sec.\ref{sec:mof}. The case of a donor-acceptor molecular crystal, such as TTF-CA (see Section\ \ref{sec:ttfca}) is an example of cooperation between  charge transfer, structural dimerization and possibly spin-Peierls transition, that ultimately leads to a paradigmatic organic ferroelectric crystal. In Sec.\ref{sec:NaLaMnWO6} we focus on recent developments in the field, as we'll discuss the case of several multiple non-polar instabilities which finally result in ferroelectricity, as shown by our ab-initio calculations for NaLaMnWO$_6$ double-perovskites. In Sec.\ref{sec:concl} we draw some conclusions and offer a perspective view of the field in the near future.


\section{General Computational Framework}
\label{sec:1}

\subsection{Density functional: technicalities}
  
All the calculations presented here have been performed 
within Density Functional Theory (DFT) 
using the Vienna-ab-initio Simulation 
Package (VASP)\cite{compu1,compu2}. The electron-ion interaction 
is described by PAW potentials\cite{compu3,compu4} using a
 plane-wave basis
set with appropriate energy cut-off. 
For Brillouin zone integrations we used the Monkhorst-Pack schemes.
We refer to the original papers (each mentioned in the appropriate section below) 
for the specific values and details 
of the computational parameters for the different compounds.

For the exchange-correlation functional, E$_{xc}$, 
we used several approximations depending on the specific problem at hand.
It is well known that there are wide classes of materials 
where density-functional methods fail not only quantitatively but also
qualitatively. Typical situations are materials with localized orbitals,
 e.g. of transition metals or rare earths. A second important class 
is comprised of the correlated organic crystals  like
 low-dimensional organic charge-transfer salts.  

The localized nature of the 3d 
electronic states  limits to the applicability 
of common density-functional methods like the
local density approximation (LDA) or generalized gradient approximation (GGA). In fact, these standard approximations 
introduce a spurious Coulomb interaction
of the electron with its own charge, i.e., the electrostatic
self-interaction is not entirely compensated. This causes
fairly large errors for localized states (e.g., Mn d states).
It tends to destabilize the orbitals and decreases their
binding energy, leading to an overdelocalization of the
charge density\cite{compu5}.
The most commonly applied Generalized-Gradient Approximation,  
the  Perdew-Burke-Ernzerhof parametrization (PBE), is often insufficient 
to treat with these problems. To overcome these failures of standard DFT approaches, 
 we make use of possible ways out.
One common approach  is the DFT+U
method\cite{compu6,compu7,compu8}, where a Hubbard-like $U$ 
term is introduced
into the DFT energy functional in order to take correlations 
partially into account. The method usually improves the 
electronic-structure description, but it suffers
from shortcomings associated with the U-dependence of
the calculated properties\cite{compu9,compu10,compu11}. 
Unfortunately, there is usually no obvious choice
 of the $U$ value to be adopted;
common choices are usually based either on experimental
input or are derived 
from constrained DFT calculations\cite{compu12}.

Another approach which is becoming 
widely used in the solid-state community 
is the use of hybrid functionals. 
Hybrid functionals go beyond the usual Kohn-Sham
formalism and fall within the generalized Kohn-Sham
realization of DFT\cite{compu13}. These functionals 
 consider  a weighted 
mixture between exchange defined in Hartree-Fock theory 
using DFT Kohn-Sham orbitals
with DFT exchange. Popular
hybrid functionals B3LYP\cite{compu14}, PBE0\cite{compu15}, and 
HSE\cite{compu16} have been constructed to give good structural, 
thermodynamic, and bonding properties of solids\cite{compu17,compu18}.
In particular, we used the HSE06 functional, 
which is particularly suitable for solid
state applications\cite{compu16,compu17,compu18}.

The ferroelectric polarization is calculated within
 the modern theory of polarization\cite{compu19,compu20}.
where one computes the difference of electric polarization,
\textit{i.e.} $\Delta$P=P$_{FE}$$-$P$_{PE}$\\=$\Delta$P$_{ion}$+
$\Delta$P$_{ele}$, 
where the subscripts $FE$, $PE$, $ion$ and $ele$ denote ferroelectric, paraelectric, 
ionic and electronic contribution, respectively. 
P$_{ion}$   is calculated by summing the
position of each ion in the unit cell times the number of its 
valence electrons. The electronic
contribution is obtained by using the Berry
 phase formalism\cite{compu20,compu21}.

 \subsection{Symmetry analysis}
 
The use of symmetry analysis in this field is very relevant, as will be evident in the following sections. We recall in this framework that,  from the symmetry point of view, a magnetic ordering breaks time--reversal symmetry, a ferroelectric ordering breaks inversion symmetry, so that {\em both} time and space inversion symmetry are absent in multiferroics (although by two different order parameters, such as for example magnetization  and polarization). Symmetry analysis can be very helpful in identifying polar ionic or electronic arrangements or in suggesting a physically-sound paraelectric structure ({\em i.e.} reasonably close to the ferroelectric structure, in terms of ionic displacements), that can be taken as reference paraelectric structure when calculating the polarization. 
Furthermore, useful tools for the calculation of the polarization 
as well as for its analysis, are  those of the 
Bilbao Crystallographic
Server\cite{compu22,compu23,compu24,compu25,compu26,compu27,compu28,compu29,compu30,compu31,compu32,compu33,compu34,compu35,compu36,compu37},
or of the ISODISTORT web site\cite{compu38}.
   
\section{Mechanisms and Materials}
\label{sec:2}
%%%and \cite{RefJ}

\subsection{Introduction: magnetic interactions and spin Hamiltonian approach}
In general, the  interactions between magnetic centers in condensed matter systems as well as in molecular systems, have a twofold orgin, one is purely magnetic and 
the other elettrostatic in nature and they 
 can be described through-space and through-bond, respectively. The former is the usual point dipolar approximation between magnetic centers, which can be safely disregarded in most cases.  The latter relies on the electrostatic interaction responsible of the 
formation of the chemical bonds. 
The states of the the interacting centres are described by a set of orbitals. As a rule of thumb for the magnetic interaction, if the single occupied orbitals are orthogonal to each other, the two spins of the electrons will be parallel to each other (ferromagnetic coupling), whereas if the orbitals have a non-zero overlap the spins will tend to orient antiparallel to each other (antiferromagnetic coupling)\cite{Kahn,Slater}.

From an historical point of view, the description of the magnetic interaction was performed by using localized magnetic orbitals or a valence-bond approach\cite{Anderson1,Anderson2,Anderson3}. Other approaches have been developed since then, which rely on the  tight-binding approaches,\cite{Kahn1,Hay} or density functional theory.\cite{Noodleman1,Noodleman2,Noodleman3,Noodleman4} 
The basis of the magnetic interaction is the antisymmetric nature of the total wavefunction, 
which shows up as an \textit{effective} exchange interaction.
The exchange interaction may occur directly (direct interaction) or through a formally diamagnetic
 ligand (super-exchange). The famous Goodenough-Kanamori rules represent a qualitatively account of
 the features of the magnetic coupling between different centers. 

It is often useful to introduce a spin Hamiltonian in order to eliminate all the orbital degree of freedoms and replace them with spin coordinates. A central approximation for such a mapping is that orbital moment is essentially quenched as it often occurs in solids, and it can be eventually 
treated as perturbation.  The spin Hamiltonian approach 
can be used for treating:
\begin{itemize}
\item Zeeman and crystal field terms for isolated ions;
\item  electron nucleus interaction terms (hyperfine interactions); 
\item interaction betweeen spin pairs.
\end{itemize}

Hereafter, we will focus on the last term. It can be represented be a spin-spin Hamiltonian, 
which can be written as 
H=$\vec{S}_{1}$ $\cdot$ \textbf{J$_{12}$} $\cdot$ $\vec{S}_{2}$  where  $\vec{S}_{1,2}$ 
are the spin operators for the magnetic center 1 and 2, respectively. \textbf{J$_{12}$} 
is a matrix which describes the interaction, which in general is not symmetric and may have a non-zero trace. It is always possible to break-down the \textbf{J$_{12}$} tensor equivalenty into three contributions:\\

-J$_{12} \vec{S}_{1} \cdot \vec{S}_{2}$ + $\vec{S}_{1} \cdot$ 
\textbf{D$_{12}$}
$\cdot \vec{S_{2}}$+\textbf{d$_{12}$} $\cdot$($\vec{S}_{1} \times \vec{S}_{2}$)\\

where \textbf{J$_{12}$}=$-$(1/3)Tr \textbf{J$_{12}$};
 D$_{12}^{\alpha,\beta}$=(1/2)(J$_{12}^{\alpha\beta}$)+
J$_{12}^{\beta\alpha}$-$\delta_{\alpha\beta}$
(1/3)Tr(\textbf{J$_{12}$}); d$_{12}$=(1/2)($J_{12}^{\beta\gamma}-J_{12}^{\gamma\beta})$ 
and $\alpha,\beta,\gamma$ are Cartesian components.  The first term in the previous formula is referred to as the isotropic which tend to keep the spins collinear;  the second as the anisotropic and it tends to orient the spins along a given orientation in space;  the third as the antisymmetric spin-spin contribution to the magnetic interaction and it tends to cant them  by 90$^{\circ}$. In many cases, the first term is the dominant one, and the other terms can be introduced as perturbation. Here, for positive J$_{12}$ we have ferromagnetic coupling, while for antiferromagnetic coupling $J_{12}$ is negative. The mechanisms at the basis of the exchange interactiona have been first introduced and discussed by Anderson, and traslated in rule-of-thumbs by Goodenough and Kanamori. 

In the following, we want to discuss the origin of the phenomenon of weak ferromagnetism and how it arises from the previous spin Hamiltonian.  The origin of weak-ferromagnetism is usaully ascribed to the so-called \\ Dzyaloshinskii-Moriya (DM) interaction.  This exchange interaction results from the interplay of the Coulomb interaction and the spin-orbit coupling in systems of low crystal symmetry. The DM interaction is an important term playing a crucial role in may magnetic systems. The weak ferromagnetism is characterized by a small net magnetic moment resulting from spin moments that nearly cancel each other. It was first observed in haematite,\\ $\alpha-$Fe$_{2}$O$_{3}$\cite{wkfm1,wkfm2}.Dzialoshinki showed that it was an intrinsic effect due to the particular symmetry properties of the crystal structure and the magnetic moments arrangments\cite{wkfm3}. Furthermore, it showed that second term and the third term 
in the previous spin Hamiltonian, \textit{i.e.} magnetocrystalline anisotropy (or anisotropic term) and  the anisotropic exchange (or antisymmetric spin-spin interaction) rispectively, can lead to a small ferromagnetic moment in an otherwise antiferromagnetic crystal. Moriya\cite{wkfm4} showed that the Dzialoshinski's explanation can be interpreted in the framework of Anderson's perturbation approach to magnetic superexchange. Furthermore, he showed that depending on the type of crystal structure either of two mechanism, magnetocrystalling anisotropy or antisymmetric exchange, can be the source for the canting of magnetic moments. For example, for $\alpha-$Fe$_{2}$O$_{3}$, it is the antisymmetric exchange that plays the dominant role whereas, in the case of NiF$_{2}$, antisymmetric exchange is ruled out in favor of the magnetocrystalling anisotropy which is here the important term giving rise to the weak ferromagnetic component. In the triangular antiferromagnetc Mn$_{3}$Sn the antisymmetric exchange contributions from different atoms cancel perfectly and can not be the reason for the observed weak ferromagnetism. Here, the magnetocrystalline anisotropy term is the source of the spin canting. Another example, is the Cu based metal-organic frameworks, 
where it can be shown that antisymmetric exchange is zero for symmetry, while the only term contributing to the spin canting is the magnetocrystalline anisotropy\cite{wkfm5}.

\subsection{Spin ordering}
\label{sec:spin}

\begin{figure}[!h]
\resizebox{0.65\columnwidth}{!}{%
 \includegraphics{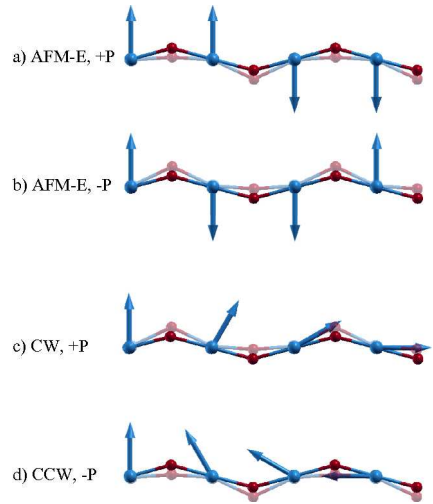}
}
%\vspace*{5cm}       % Give the correct figure height in cm
\\
\caption{Examples of polar spin configurations (shown in the MnO$_2$ basal plane of $Pnma$ manganites): a) and b)  collinear cases (such as AFM-E in HoMnO$_3$) with opposite polarization, induced by flipping the direction of two spins. c) and d)  spiral cases (such as TbMnO$_3$) with opposite polarization, induced by the different clockwise (CW) vs counterclockwise (CCW) rotation of spins in the vertical plane. The consequent exchange-strictive mechanisms are also shown: semi-transparent oxygens schematically show the displaced atoms (exaggerated for clarity) occurring upon spin ordering: In a) and d) the  spin configuration is such that oxygens move ``down", whereas in   b) and c)  oxygens move ``up".  }
\label{fig:spins}       % Give a unique label
\end{figure}

In recent years, people have focused mainly on two different mechanisms of magnetically-induced ferroelectric polarization, P, driven by either symmetric ${\bf S_i \cdot S_j}$ Heisenberg (H) 
exchange\cite{prlslv,sergienko2007} or antisymmetric  ${\bf S_i}$ x ${\bf S_j}$ spin--orbit--like Dzyalo\-shin\-skii-Moriya (DM) 
exchange\cite{knb2005,sergienko2006,mostovoy2006}. 
The prototypical materials taken as representative examples of 
the two mechanisms are  by far HoMnO$_3$ and\\ TbMnO$_3$, 
respectively: {\em i}) the first manganite shows \\a collinear $\uparrow-\uparrow-\downarrow-\downarrow$  spin configuration along [101] and [10-1] directions in the $Pnma$ setting ; alternatively,  this peculiar AFM ordering can be seen as zig-zag ferromagnetic chains antiferromagnetically coupled in the $ac$ MnO$_2$ plane . This so--called AFM-E spin order (which is stable when the A-site ionic radii is small, such as Tm, Lu, Yb rare-earth cations in ortho-manganites) clearly lacks inversion symmetry (see Figure\ \ref{fig:spins} a).  From the microscopic point of view, the polarization comes from the inequivalency between the oxygens bonded to two Mn with parallel spins, with respect to those linked to Mn with antiparallel spins, as discussed in detail in 
Refs.\cite{review4,prlslv,wannier.yamauchi} {\em ii}) the second manganite, TbMnO$_3$ shows as ground-state a spin cycloidal spiral 
in the $bc$ plane. In that case, non-collinear spins induce polarization,  as 
predicted by the formula {\bf $P\propto e_n \times Q $},	where {\bf $Q$} is the spiral wave-vector parallel to the chain direction
and {\bf $e_n \propto S_n \times S_{n+1}$} is the spin-rotation axis [see Figure\ \ref{fig:spins} c) and d)] The vector product   {\bf $S_n \times S_{n+1}$} can be shown to be proportional to the spin current ${\bf j_s}$,
in turn linked, via a vector-potential relation\cite{knb2005}, 
to the Dzyaloshinskii-Moriya interaction, so that  
the mechanism of ferroelectricity in spin-spirals is
 generally labeled as ``spin--current--induced"\cite{knb2005}.

In the context of spin-induced ferroelectricity, it is interesting to discuss how ``switching" of polarization occurs. We recall that, in standard ferroelectrics,  ionic displacements occur in equal magnitudes and opposite directions (with reference to the paraelectric state) when reaching the ``+P" or ``-P" state. However, in spin-induced ferroelectricity, since polarization is determined primarily by the spin ordering and not by ionic displacements, one expects some changes to occur in the arrangement of magnetic moments. For example, in AFM-E ortho-manganites, switching occurs when  half of the spins in the unit cells have their direction flipped by 180$^\circ$, so that  all oxygens that were previously connected to parallel spins now become oxygens being bonded to antiparallel spins (and viceversa, cfr 
Figure\ \ref{fig:spins} a) and b)). When focusing on spiral TbMnO$_3$, switching occurs when reversing the vector spin chirality, {\em i.e.} when reversing a clockwise spiral into an anticlockwise spiral, as shown in
 Figure\ \ref{fig:spins} c) and d).

What is important to emphasize for spin-driven ferroelectricity is again the difference between the {\em electronic} and {\em ionic} contributions to polarization: the first arises from the polar rearrangement of the electronic charge, it is present even with ions arranged in a centrosymmetric  way and was shown to be relevant in both the spiral and AFM-E cases; the second is mainly referred to as exchange-striction (symmetric for the Heisenberg case and antisymmetric for the DM case), meaning that the ions move, when the spin ordering is established, to gain energy  from exchange terms
 (see Figure\ \ref{fig:spins} for a schematic representation). Which of the two - electronic vs ionic - contributions is more relevant actually depends pretty much on the considered system: from first-principles calculations, the two were found to be almost the same in 
AFM-E HoMnO$_3$\cite{prlslv}, whereas the ionic was estimated to be much larger in spiral
 TbMnO$_3$\cite{malash}.

When comparing the two DM and H mechanisms, one expects  that, being the former DM-induced polarization driven by relativistic effects which are not so strong in 3d-based materials, the DM-related polarization should be weaker than the H-induced one.
Indeed, in a variety of materials (ranging from nickelates, such 
as LuNiO$_3$\cite{nickel}, to manganites, such as HoMnO$_3$\cite{prlslv},
 or sulfides, such as
 Cu$_2$MnSnS$_4$\cite{tets}), we have shown that 
the size of P induced by the relativistic DM interaction
 (occurring in spin-spiral-based oxides) is much lower than 
that caused by the H interaction. %Furthermore, our predictions were recently confirmed experimentally, proving that polarization up to $\leq$ 1 $\mu$C/cm$^2$ can be achieved.
Related to this, a comment is in order: 
it was very controversial how large the polarization actually
 was in AFM-E like rare-earth ortho-manganites, a paradigmatic 
case of Heisenberg-driven ferroelectricity. 
Earlier experiments from Lorenz 	 
{\em et al.}\cite{lorenz} reported ferroelectricity on polycrystalline samples of ortho- HoMnO$_3$ of the order of 10$^{-3}$ $\mu$C/cm$^2$, three orders of magnitude smaller than what predicted by our DFT calculations. Recently, however, the group of Tokura, \cite{ishiwata} by means of advanced growth techniques and measurements, focused on several rare-earth ortho-manganites (RMnO$_3$, R = Ho, Lu, Eu$_x$Y$_{1-x}$) estimated the genuine values of P 
(about 0.5 $\mu$C/m$^2$) in the E-type phase, 
which is more than 10 times as large as that of
 the $bc$ cycloidal phase. Slightly earlier, a
 theoretical model was reported in the context of electromagnon excitations in
 RMnO$_3$ \cite{valdez}. One of the outcome was the estimate of the polarization in E-type manganites based on optical absorption data measured for TbMnO$_3$ in the spiral-phase: P was found to be of the order of 1 $\mu$C/cm$^2$. Moreover, 
Pomyakushin \textit{et al.} \cite{Pom} have reported the polarization of about 0.15 $\mu$C/cm$^2$ for E-type TmMnO$_3$. There is therefore now a growing consensus on the possibility of breaking inversion symmetry (therefore paving the way to improper ferroelectricity) via the symmetric magnetic exchange (Heisenberg-like) in collinear frustrated systems, in addition to the (well consolidated) analogous effect in the antisymmetric counterpart (Dzyaloshinskii-Moriya-like) of magnetic exchange. 

%The study on Cu$_2$MnSnS$_4$, where Mn ions are located rather far apart (as fourth nearest neighbors, see below for details), shows that, in those cases where a polar AFM order induces ferroelectricity, the size of the ferroelectric polarization is related � as expected � to the Mn-Mn interaction. For example, in our study of (undoped) AFM-E
%manganites (where ferroelectricity is driven by a polar AFM order), the O ion displaces quite differently, depending on whether the two Mn to which it is bonded are parallel or antiparallel (i.e. the Mn-O-Mn angle for ferromagnetically-coupled Mn is much larger than for AFM-coupled ions). As such, the displacements of the O sublattice in E-type occur in a non-centrosymmetric way and largely contribute to polarization. The stronger the interaction, the larger the consequent �exchange- striction� that might strengthen the permanent dipolar moment. On the other hand, in the Cu2MnSnS4 sulfide case (see below), the S atoms - once the polar AFM ordering on the Mn ions is imposed - are displaced little from their original centrosymmetric positions. As a result, the ionic contribution to polarization is negligible and the ferroelectricity has mostly an electronic character. A conclusion that can be drawn is therefore that, in those cases where the ionic contribution to polarization would strengthen the electronic one, it is better to choose a �magnetically-concentrated� material, i.e. with a �dense� magnetic sublattice. In fact, the latter could induce a substantial ionic contribution to polarization coming from exchange-strictive effects.

 In what follows, we'll discuss two peculiar cases in which the Heisenberg exchange striction is at play to induce ferroelectricity, according to a microscopic mechanism based on the up-up-down-down spin chain: a rare-earth orthoferrite and Cd-based vanadate. Despite the similarity with the prototypical HoMnO$_3$, both systems show some peculiarities: in the first case, it is the interaction between 4$f$ and 3$d$ spins, {\em a-priori}  not expected to be very strong, that causes a sizeable {\bf P}, whereas in the second case, it is the peculiar spinel coordination - and related oxygen arrangement - which makes the mechanism underlying ferroelectricity more intriguing. 

\subsubsection {$f-d$ coupling in DyFeO$_3$}
\label{sec:dfo}

\begin{figure}[!h]
\centerline{
\resizebox{1.0\columnwidth}{!}{%
  \includegraphics{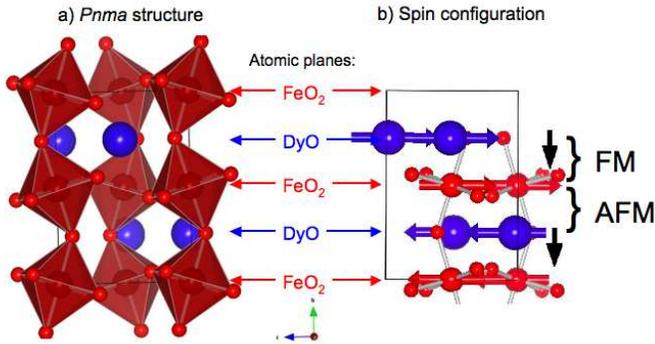}}}
 
\vspace*{2.4cm}       % Give the correct figure height in cm
\caption{Orthorhombic DyFeO$_3$: a) crystal structure (FeO$_6$ octahedra and atomic planes highlighted); b) spin configuration: blue (red) arrows denote Dy (Fe) spins. Black bold arrows denote the displacement (with respect to the non-magnetic paraelectric  structure) of FeO$_2$ planes, due to exchange-striction.}
\label{fig:dfo}       % Give a unique label
\end{figure}

DyFeO$_{3}$ has been studied since a long time for its interesting physical
 properties\cite{dfo1,dfo2,dfo3,dfo4,dfo5,dfo6,dfo7,dfo8,dfo9,dfo11,dfo13,dfo14,dfo15}.
Dysprosium-Orthoferrite, DyFeO$_3$ (cfr Figure\ \ref{fig:dfo} a),
 was suggested	from \\experiments\cite{tokunaga} to
 show, below the ordering temperature of Dy and upon application 
of a magnetic field parallel to the $c$-axis and larger than a 
specified critical magnetic field, a MF phase, with  weak	
ferromagnetism	and ferroelectricity (M$\sim$0.5 $\mu_B$ per formula-unit 
and P $\sim$ 0.2 $\mu$C/cm$^2$). By performing accurate DFT calculations
 (see details in Ref.\cite{njp}), we focused on the mechanism at
 the basis of the polar behavior, neglecting the weak magnetic moment 
that most likely arises from a Fe spin-canting (not directly related
 to ferroelectricity) and considering a simpler collinear spin arrangement. We showed that 4$f$ electrons, more-often-than-not neglected in modelling studies, play an unexpected and important role in stabilizing the magnetic-field-induced ferroelectric state of DyFeO$_3$. Indeed, we found the FE polarization to be mainly driven by an exchange-strictive mechanism, working between adjacent spin-polarized Fe and Dy layers arranged in a Fe$\uparrow$-Dy$\uparrow$-Fe$\downarrow$-Dy$\downarrow$ fashion 
(see Figure\ \ref{fig:dfo} b): DyO atomic planes move towards FeO$_2$ ionic planes so as to maximize the energy gain coming from a Fe-3$d$/Dy-4$f$ ferromagnetic coupling. 
Here, at variance with AFM-E HoMnO$_3$ where  
the $\uparrow\uparrow\downarrow\downarrow$ spins are all equivalent,
 we have an intrinsic inequivalency between Fe-$d$ and Dy-$f$ spins, 
so that Fe$\uparrow$-Dy$\uparrow$-Fe$\downarrow$-Dy$\downarrow$ 
chains ``dimerize" and give rise to polarization. Indeed, the distortions 
lead to an alternate short-long-short-long interlayer distance between
 DyO and FeO$_2$ planes, with $d_{FM}$ = 1.898\ \AA $\:$ and 
$d_{AFM}$ = 1.910\ \AA, to be compared with $d_{ideal}$ = 1.904\ \AA $\:$ in
 the unrelaxed non-spin-polarized case (cfr Figure\ \ref{fig:dfo}).
Indeed, we were able to identify two degenerate and switchable 
polar states, {\em i.e.} characterized by a sign-reversal of the
 FE polarization ($\pm$P) and connected by a relative rotation of 
the direction of Dy spins (with respect to Fe spins). The estimated 
FE polarization, in good agreement with experiments\cite{tokunaga}.
 shows an unexpectedly large magnitude 
($\sim$0.1-0.2 $\mu$C/cm$^2$).

So far we have discussed  the exchange striction mechanism in the 
  Heisenberg framework. Obviously, 
this coupling must have a microscopic
explanation in terms of   interaction between 
orbitals. In order to unveil the microscopic origin of the polarization, 
a careful analysis of atomic displacements as well as 
of electronic structure is needed.

Let's consider the virtual paraelectric phase where the ions are locked at 
a centrosymmetric (CS) position ($Pnma$ setting, point group $D_{2h}$). A suitable spin configuration which does not break the inversion symmetry shows the Dy spin intralayer FM coupled, but rotated with respect to the Fe spins by 90$^{\circ}$. The FE$_{1}$ ($+P$) or FE$_{2}$ ($-P$)  state can be obtained by progressively rotating the Dy spins in the $ac$ plane, counter-clockwise for FE$_{1}$ or clockwise for FE$_{2}$, when viewed from the positive direction 
of the $b$ axis\cite{njp}. In this way, one recovers the AFM-A spin configuration, {\em i.e.} ferromagnetic planes antiferromagnetically coupled along the $b$ axis.

From symmetry point of view, the configuration with ortogonal spins has the \textit{magnetic} space group $P212121$ (N. 19.1.119), and therefore the space group for the nuclear
sites is $P212121$ (N. 19). This means that the magnetic ordering breaks
all the symmetry operations containing  time inversion, plus the inversion
centre and all mirror planes. However, we remark that, although the magnetic ordering
breaks the inversion centre, the resulting space group symmetry is
\textit{non-polar}. It should be noted that the nuclei are not constrained by this
symmetry to stay in the ideal  $Pnma$ configuration that we  used,
and can in principle relax to a more general arrangement compatible
with the $P212121$ space group, by means of a non-polar distortion. Also
the magnetic symmetry does not force the magnetic moments to be
strictly orthogonal. Both  Fe and Dy magnetic moments could have
some antiferromagnetic components along some of the other axes 
(with different
sign correlations among sites)\cite{ManuelComments1}. 
In our work, we have disregarded these (presumably small)
deviations from  collinear spin configuration. 

The configuration with collinear spins has the \textit{magnetic} space group
$Pn21a$ (N. 33.1.226 in non-standard setting), and therefore the space
group for the nuclear sites is $Pn21a$ (N. 33 in non standard setting).
This means that the magnetic ordering breaks all the symmetry operations containing
 time inversion, plus the inversion centre, the binary axes along $x$ 
and $z$, and the mirror plane perpendicular to $y$. This magnetic ordering
breaks the inversion centre as in the previous case, but now the 
resulting space group symmetry is indeed \textit{polar} along the $b$-axis. Again, the
nuclei are not constrained to stay in the ideal $Pnma$ configuration
that we  used, and could relax to a more general
arrangement compatible with the space group $Pn21a$, through a
distortion which will be polar along $y$. So in principle, it is quite
similar to the first case above, the difference being that the
possible distortions relaxing the nuclear positions will be polar, and
therefore can yield some macroscopic polarization along the $b$-axis,
while in the previous case the possible relaxations of the nuclear
structure are necessarily non-polar. As in the previous case also, the
magnetic symmetry of the configuration does not force the magnetic
moments to be strictly collinear, and some additional antiferromagnetic
arrangements of other additional components of the magnetic moments
(with different sign relations among the sites)  are  allowed by symmetry,
and,  however small, they will in principle be present in a fully
 relaxed structure\cite{ManuelComments2}.
 Also in this case, we have not considered these spin components.

Let's focus on equatorial oxygens, O$_{eq}$, which occupy the 
$8d$ Wyckoff positions (WPs). 
In the FE phase, 
when the symmetry is lowered to C$_{2v}$, the
O$_{eq}$ become \textit{inequivalent} and  a WP splitting 
$8d\rightarrow4a+4a$ shows up.  Inspection into the 
local spin configuration around O$_{eq}$s explains the reason of this 
inequivalency: the O$_{eq}$ sandwiched by  FM coupled
Fe and Dy layers,  have  two $\uparrow$Fe and
 two $\uparrow$Dy atoms as nearest neighbors 
(we call them as  O$_{eq}^{\uparrow,\uparrow}$); 
when sandwiched by Fe and Dy layers AFM coupled, 
they have   two $\uparrow$Fe and two $\downarrow$Dy 
atoms as nearest neighbors
(labelled as   O$_{eq}^{\uparrow,\downarrow}$). 
The inequivalency due to the local spin environment is confirmed by the 
following computational experiment: if 
we impose the FE$_{1}$ (or FE$_{2}$) spin configuration 
on top of the centrosymmetric (CS) ionic structure, 
O$_{eq}^{\uparrow,\uparrow}$ and  O$_{eq}^{\uparrow,\downarrow}$ 
become  \textit{inequivalent}: 
O$_{eq}^{\uparrow,\uparrow}$ 
has $\pm$0.194 $\mu_{B}$ and  O$_{eq}^{\uparrow,\downarrow}$ 
has $\pm$0.207 $\mu_{B}$ 
as induced spin moment.  This time, however, no WP splitting is involved, 
since the ions are frozen in CS positions.
To rule out any numerical artifact on this small difference, 
we impose the PE spin configuration on top of the CS ionic structure. 
In this case, all O$_{eq}$s carry induced spin moments of 
exactly the same magnitude, becoming equivalent again.
This leads to the conclusion that the change of spin state
in going from PE to FE$_{1}$ must be the source of the inequivalency 
of  O$_{eq}$s, and, eventually, it should be strongly correlated
 to the presence of ferroelectricity. If so, the ferroelectric state in our toy-model is spin-induced. To support this conclusion we note that 
in the PE spin configuration on top of the CS positions, P$_{tot}$=0 while in FE$_{1}$, 
P$_{tot}$ is different  from zero, \textit{even when the ions are at CS positions}.

In passing we note that all oxygens remain equivalent 
when  the Dy-$f$ electrons are not treated (as done so far) as valence electrons, but they are treated as ``frozen" in the core. 
In this computational experiment,
 the Dy atoms  lose their spins and  
O$_{eq}^{\uparrow,\uparrow}$$\rightarrow$
O$_{eq}^{\uparrow,nospin}$ and 
 O$_{eq}^{\uparrow,\downarrow}$$\rightarrow$
O$_{eq}^{\downarrow,nospin}$: O$_{eq}^{\uparrow,nospin}$ is equivalent to 
O$_{eq}^{\downarrow,nospin}$ since here we are neglecting the spin-orbit coupling.
We are, therefore, led to the conclusion that a signature of the FE instability is the spin-induced inequivalency of O$_{eq}$, which, in turn, must be correlated to Dy-$f$ states, 
which carry the Dy spins. An analysis of the  symmetry 
breaking 
distortions\cite{compu22,compu23,compu24,compu25,compu26,compu27,compu28,compu29,compu30,compu31,compu32,compu33,compu34,compu35,compu36,compu37,compu38}
sheds further light into the microscopic mechanism.
The mode decomposition confirms that a \textit{polar} mode is involved, 
called $GM4-$. The corresponding pattern of 
atomic displacements (not shown here, for details see 
Ref.\cite{njp})  with respect to the CS structure highlights the subtle 
inequivalency of O$_{eq}$s: 
O$_{eq}^{\uparrow,\uparrow}$ (O$_{eq}^{\uparrow,\downarrow}$) 
move in such a way to \textit{decrease}  (\textit{increase}) 
the distance to its neighbor Dy atom. 
For O$_{eq}^{\uparrow,\uparrow}$, d$_{Dy-O}$ is 2.478 \AA; 
for  O$_{eq}^{\uparrow,\downarrow}$, d$_{Dy-O}$ is 2.496 \AA\ 
(the corresponding distance in the PE phase is 2.487 \AA). This would  suggest
that a weak bonding interaction is active between the FM layers, in turn 
responsible for the changes in the distances the magnetic layers,
{\em i.e.} for the dimerization and the rising of the polarization.

A useful tool to study tiny differences in bonding interaction
in solid state systems is the 
electron localization function (ELF)\cite{Savin1,Savin2}.
The electron localization function was introduced by
Becke and Edgecombe as a measure of the probability of
finding an electron in the neighborhood of another electron
with the same spin.\cite{Savin1,Savin2} ELF is thus a measure of the Pauli
repulsion, which is active when same spin-electron wavefunctions start to overlap causing 
a repulsion due to antisymmetrization postulate. The ELF values lie by definition between zero and one. Values
are close to 1, if in the vicinity of one electron no
other electron with the same spin may be found,
 for instance as occurring
in \textit{bonding pairs} or lone pairs\cite{Savin2}.

We here look for  a \textit{signature} that two ferromagnetic 
layers are interacting
through exchange striction. In terms of chemical bond picture,
there should be a ``bond'' formation between the two layers.
In our case, we want to show that this bond formation is basically due
to the presence of $f$-electrons of Dy through a direct 
(or indirect) interaction. Indeed, we have seen that if
we consider the $f$ electrons in the core 
({\em i.e.} using the VASP code, we   use 
the Dy$\textrm{\_3}$ potential),
we don't observe any exchange striction effect.
If we consider  them in the valence 
(and so let them eventually interact), 
we do have an exchange striction effect,
{\em i.e.} the two ferromagnetic sheets
 approach each other. Although the effect is small, it disappears
completely when using Dy$\textrm{\_3}$. This is clearly confirmed by
 the following computational
experiment: starting from the FE$_{1}$ ionic structure, 
we freeze the $f$ electrons 
in the core (using Dy$\textrm{\_3}$ POTCAR file within VASP). 
Obviously the Dy atoms are not spin-polarized
in this case. We let the system to relax to its new electronic and ionic 
ground state. Not surprisingly, we found that the system 
relax to a non-polar state (PE state). Viceversa, starting from the PE ionic structure, 
and treating the $f$ electrons in the valence, the FE state is stabilized.
In summary, when considering the $f$ electrons as valence states, 
the PE state becomes unstable, 
the D$_{2h}$  point group symmetry is spontaneously broken
 to C$_{2v}$ and the system evolves towards a stable and 
polar state. If the $f$ electrons are removed from the 
valence and frozen in the core,
 the PE state remains stable. This unambiguously confirms 
that $f$ states are a necessary ingredient for
 ferroelectricity in DyFeO$_{3}$.
So, coming back to our ELF function,  
we consider the \textit{difference}  in ELF (DELF)
between the situation when $f$ electrons are in the
valence (which stabilizes the FE state, as previously observed)
and when they are frozen in the core (thus stabilizing the PE state),
 for the same ionic configuration (for instance the CS), \textit{i.e.} 
DELF($\vec{r}$)=ELF$_{f_{val}}$($\vec{r}$)-ELF$_{f_{core}}$($\vec{r}$).
The physical interpretation is as follows: positive values 
of DELF show up in regions where the electron localization is higher, 
\textit{i.e.} the bonding between FM layers is strengthened. 
It can be shown\cite{njp} that  a \textit{positive} 
isosurface  of DELF 
projected into the $ab$ plane is  
mainly localized between FM layers and, more specifically,
 in the region between Dy and O$_{eq}^{\uparrow,\uparrow}$. 
This points therefore to a bonding interaction between FM
layers mediated by O$_{eq}^{\uparrow,\uparrow}$.

Our electronic structure analysis interpret this finding as an 
  efficient mediation of O-2$p$ and
 Dy-$d$ between the relevant Dy-4$f$ and Fe-3$d$ electrons. For 
details see Ref.\cite{njp}. These results pave therefore the way to the interaction between $f$ and 
$d$ electronic states as an additional degree of freedom to tailor 
ferroelectric and magnetic properties in multiferroic compounds. 
Additionally, the fact that the Dy and Fe interaction is mediated 
by O-$2p$ states suggests possible 
routes to tailor the FE polarization. For instance, compressive or 
tensile strain along the polar
axis might change the  octahedral tilting, favoring or disfavoring 
the interaction via the
intermediate O states, eventually leading to a  change in the FE 
polarization.
From the methodological point of view, Dy-ferrite constituted a 
nice benchmark for the theoretical treatment of 4$f$ electrons, 
usually a hard task within ab-initio approaches. Our results were 
shown to be robust with respect to the different state-of-the-art 
computational schemes used for $d$ and $f$ localized states, such 
as the DFT+U method, the Heyd-Scuseria-Ernherof (HSE) hybrid 
functional, and the GW approach\cite{njp}.

\subsubsection {$\uparrow\uparrow\downarrow\downarrow$ spin arrangement in CdV$_2$O$_4$}
\label{sec:cvo}

%%%\label{sec:cvo}

Spinels exhibit several unusual features, which make them an emerging class of materials in several fields, including magnetoelectricity and multiferroicity. Most oxide-based spinels contain trivalent and divalent cations and have the general formula 
A$^{+2}$B$^{+3}$$_{2}$O$_{4}$. The structure type is that of the  MgAl$_{2}$O$_{4}$ mineral, which is cubic and contains eight formula units.  Oxygen ions show  a cubic close-packed arrangement and the cations occupy both octahedral and tetrahedral sites. When the tetrahedral sites are occupied  by divalent cations only and the octahedral sites by trivalent ions only, the structure is termed a \textit{normal} spinel.

The series of Vanadium oxide spinels with A = Cd, Mn, Zn or
Mg and B=V is particularly interesting, since it 
approaches a Mott transition when
the V-V distance is reduced sufficiently, either by  applying pressure
or by changing the size of the A cation\cite{cvo1,cvo2}.
ZnV$_{2}$O$_{4}$ is the  member of the series which is closest to the metallic state.
The ground state of the spinel compounds has stimulated
an intense theoretical research in the last 
few years\cite{cvo3,cvo4,cvo5,cvo6,cvo7}.
A tetragonal distortion induces the t$_{2g}$ levels to split
into a lower d$_{xy}$ level and a  twofold degenerate d$_{xz}$ and d$_{yz}$
level. It is clear that the first electron of V$^{+3}$ ($d^2$) occupies
 the d$_{xy}$ level, whereas the second one is located in a
combination of the other t$_{2g}$-orbitals (d$_{xz}$ and d$_{yz}$).

Three main models have been proposed to describe the  d$_{xz}$ and d$_{yz}$ occupations:
the ``real" orbital--ordered model, where the ground state
consists of alternating occupation of the d$_{yz}$ or d$_{xz}$ orbital in
adjacent layers along the $c$ axis; the ``complex" orbital order
model, where the second electron occupies the complex orbital
(d$_{yz}$$\pm$d$_{xz}$) which has an unquenched value of the orbital
angular momentum;  a third model  takes into
account the proximity of ZnV$_{2}$O$_{4}$ to the itinerant-electron
boundary, showing a dimerization along the V-V chains,
which is described by the formation of homopolar molecular
V-V bonds, characterized by a partial electronic delocalization
(the orbital wave functions in this case would be a
real combination of orbitals (d$_{yz}$$\pm$d$_{xz}$) with no net orbital
moment). As far as the magnetic structure is concerned, chains in the $xy$ ($ab$) plane show
an  AFM  ordering, while chains in the $xz$ and $yz$ plane have $\uparrow\downarrow\uparrow\downarrow$
spins. The overall spin structure is thus \textit{collinear}. 
It must be noted that magnetic A sites like Mn or Fe have a much more complicated magnetic structure.  The multiferroic behaviour 
in the spinel class of materials is very rare. 
The only few exceptions are CoCr$_{2}$O$_{4}$, 
HgCr$_{2}$S$_{4}$ or CdCr$_{2}$S$_{4}$. Usually, 
a more complex magnetic structure is involved there, such as  spiral magnetism.

A multiferroic
behaviour was recently reported  in a ternary spinel, CdV$_{2}$O$_{4}$ (hereafter called CVO)
with colli\-near antiferromagnetic ground state, 
where the ferroelectricity arises from a local-exchange striction mechanism.
This shows that not only spiral magnetism can give rise to polarization in spinels,
but also collinear structures through exchange striction, thus broadening  
the class of  multiferroic systems by including this class of oxides.
\begin{figure}[!h]
\resizebox{0.45\textwidth}{!}{\includegraphics{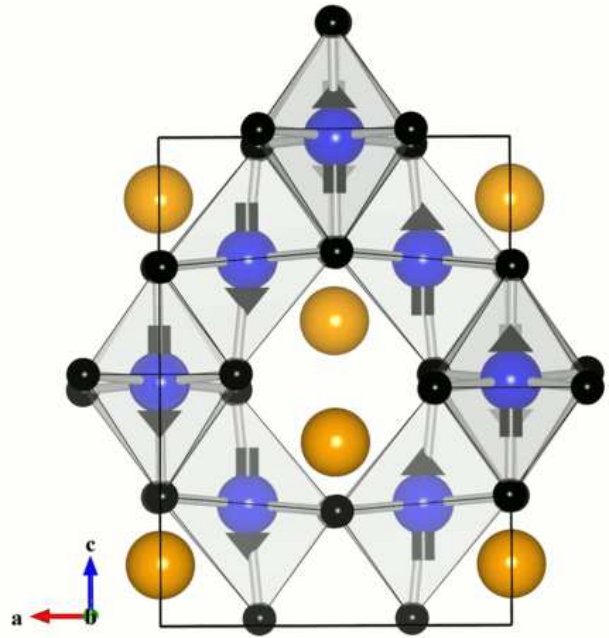}}
\caption{Perspective view of the spin and crystal structure of the CdV$_{2}$O$_{4}$.} 
%The 
%$\uparrow\downarrow\uparrow\downarrow$ spin chain is highlighted by the dashed-line.}
\label{cvo_fig1}
\end{figure}  

In Figure\ \ref{cvo_fig1} we show a perspective view of the unit cell of CVO, along with the spin structure.
It is easy to recognize the $\uparrow\uparrow\downarrow\downarrow$ chains. In this particular 
case, ab-initio calculations proved to be very useful to highlight the microscopic mechanism leading to a finite $P$ especially through a series of computational experiments and a trend study  as a function of the $U$ parameter. 
Our study can be summarized as follows:\\
\begin{itemize}
\item We started from the centric $I4_{1}/amd$ space group and we imposed 
 $\uparrow\uparrow\uparrow\uparrow$ spins. In this initial ionic configuration, there was no V-V dimerization. Then, for any $U$ between 0 and 8 eV, 
we let the ionic degrees of freedoms  relax.
What about  the final ionic and electronic ground states? 
We found: i) no V-V dimerization; ii) no inversion symmetry-breaking and thus, no polarization (for $U$ sufficiently large to keep the system insulating, {\em i.e.} larger than $U$=4 eV);
\item  Starting from the centric $I4_{1}/amd$ symmetry, we then imposed $\uparrow\uparrow\downarrow\downarrow$ spins along the [101] and [011] directions. After ionic relaxations, we found: i) the formation of short (S) and long (L) bonds between $\uparrow\uparrow$ and $\uparrow\downarrow$ spins, respectively; ii) inversion symmetry breaking and appearance of a  finite polarization. Although we found it difficult to unambiguously extract the final symmetry group of the relaxed structure, due to numerical noise, a compatible symmetry group may be 
the space group 80,  C4-6,  I4$_{1}$ (The threshold on the atomic position for the symmetry check has been fixed to  0.001 \AA)
\end{itemize}
\begin{figure}[!h]
\resizebox{0.65\textwidth}{!}{\includegraphics{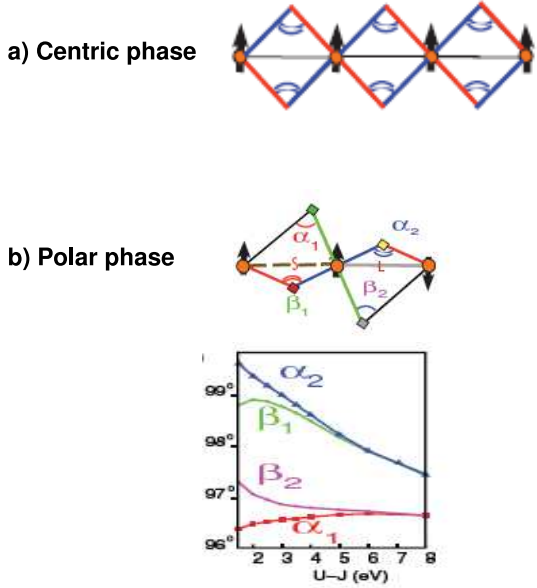}}
\caption{Sketch of the V spin chain in the  (a) centric and in the  (b) polar phase. Lower panel:
Relevant V-O-V angles (as defined in b)) along the the spin chain as a function of the effective Coulomb parameter (U-J) within the DFT+U formalism. 
(Adapted from Fig.\ 1 of Ref.\ \cite{cvo16}.)}
\label{cvo_fig2}
\end{figure}

Further inspection in the  mechanism for ferroelectricity, shows that:
the  $\uparrow\uparrow\downarrow\downarrow$ spin order, imposed onto the centric $I4_{1}/amd$ 
structure, gives rise to an electronic instability that ultimately results in {\em i)} a V-V dimerization and {\em ii)} formation of short and long V-O bonds, 
compatible with a staggered $xz$,$yz$ orbital ordering. It is worthwhile to note that this electronic driving force towards {\em i)} and {\em ii)} shows up already before performing ionic relaxations: the two $\uparrow$ ($\downarrow$) V sites are inequivalent and such inequivalency, upon relaxations, drives the V-V dimerization; the two oxygens bonded to $\uparrow$,$\uparrow$ V (or $\downarrow$,$\downarrow$ V) are inequivalent, in turn giving rise, upon ionic relaxations, to a weakly staggered orbital ordering. We believe that both effects, {\em i.e.} dimerization and orbital ordering, cooperate to induce polarization. 
In Figure\ \ref{cvo_fig2} we show the  centric phase with a FM spin chain, and the polar phase
with an $\uparrow\uparrow\downarrow\downarrow$ spin chain. 
In the centric phase, 
all V-O-V angles are equivalent along the chain, see Figure\ \ref{cvo_fig2} (a).
Note, however, that V-O distances are slightly different, due
to the peculiar coordination of the spinel structure: Each O
is an ``apical" one with respect to one V ion and a ``planar"
one with respect to the other neighboring V ion. 
This is in principle compatible with the presence of partial orbital ordering, even in the FM spin chain.
 As expected from the centrosymmetric
space group, no polarization is found from our calculations
for this case. In the polar phase, see Figure\ \ref{cvo_fig2} (b),
(i) the angles $\alpha_{1}$ and $\beta_{2}$ ($\alpha_{2}$ and $\beta_{1}$) become
inequivalent due to the formation of short and
long V-V bonds; (ii) $\alpha_{1}$ and $\beta_{1}$ become different. The long
V-O bonds are   compatible with a weakly staggered $xz$,$yz$ orbital ordering.
As a result, local dipole moments, originating from
the inequivalency of oxygens, appear due to different $\alpha_{1}$ and $\beta_{2}$ ($\alpha_{2}$ and $\beta_{1}$)
angles; since the dipoles do not compensate,   we observe a net $P$ in the unit cell. 
Further details can be found in Ref.\cite{cvo16}.

\subsection{Charge ordering}
\label{sec:CO}
\begin{figure}[!h]
\hspace{2cm}
\resizebox{.8\columnwidth}{!}{%
  \includegraphics{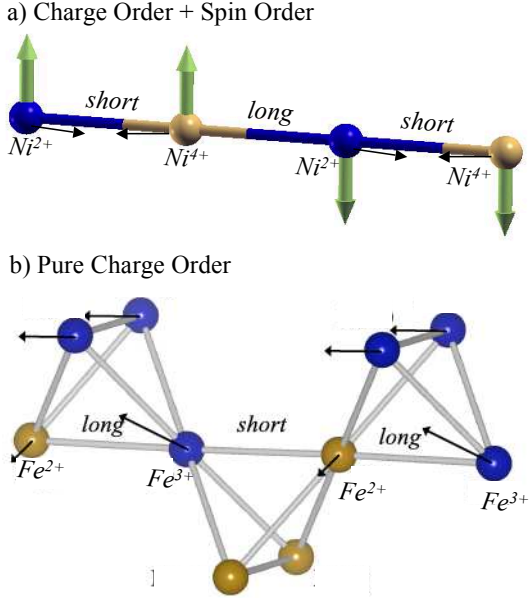}}
\vspace*{2cm}       % Give the correct figure height in cm
\caption{a) Coexistence of charge and spin orders, whose cooperation breaks inversion symmetry and gives rise to ferroelectricity. Shown is the case of RNiO$_3$ (R = rare earth), where Ni - nominally trivalent - charge-disproportionates  into Ni$^{2+}$ and Ni$^{4+}$ and  magnetically orders in an $\uparrow-\uparrow-\downarrow-\downarrow$ fashion along the [111] pseudo-cubic direction. Due to the Ni spin and charge inequivalence, a dimerization is induced along [111] and polarization arises. b) Pure charge-ordering giving rise to polarization. Shown is the case  of Fe in magnetite along the $b$ direction, where a dimerized chain of Fe$^{2+}$-Fe$^{3+}$ forms, due to a complex interplay between Coulomb repulsion in the Fe tetrahedral network, entropy, electron-phonon interaction, etc. Black thin arrows denote  Fe displacements with respect to the centrosymmetric configuration.}
\label{fig:CO}       % Give a unique label
\end{figure}

Transition metal oxides often show correlated electrons, which, under certain conditions (due to a complex interplay among Coulomb repulsion,  electron-phonon interaction,  Jahn-Teller effects, etc), can lead to electronic charges being localized on different ionic sites (the latter belonging to the same chemical species) in an ordered fashion: 
this phenomenon is labeled  as charge-disproportionation or 
Charge ordering (CO)  and it is a (first- or second-order) phase transition, with well defined critical temperatures. 
Charge ordering (CO) was proposed as a phenomenon that can induce ferroelectricity in those cases where, similar to spin ordering, the symmetry of the CO pattern below the critical ordering temperature lacks inversion 
symmetry\cite{jeroen}. Several examples were put forward:
Polar CO was identified in Fe$^{2+}$/Fe$^{3+}$ occurring 
in magnetite\cite{alexe,yamauchi.prb,tets1}, Mn$^{3+}$/Mn$^{4+}$ in half-doped 
manganites\cite{GGhalfdoped,efremov}, Ni$^{2+}$/Ni$^{4+}$ in rare-earth 
nickelates\cite{nickel}. Ab--initio calculations showed that a polar CO can lead to potentially large  polarization (of the order of few $\mu$C/cm$^2$), so that it can be considered as an efficient mechanism in the context of electronic ferroelectricity.  
Among the mentioned systems, however, one has to make an important distinction as for the driving mechanisms:
there are cases, such as   nickelates and manganites, where multiferroicity shows up when {\em both spin and charge orders} occur. It is actually their combination that drives ferroelectricity 
(see Figure\ \ref{fig:CO} a). In this situation, magnetism has a relevant role in the development of the dipolar order (actually, magnetism and ferroelectricity share the same physical origin) and a large ME coupling is therefore expected. In other cases, such as magnetite, %on the other hand, where CO by itself drives ferroelectricity (see Fig.\ref{fig:CO} b), 
deeply discussed in the following paragraph, magnetism seems not to be involved in the ferroelectric transition. %(Fe$_3$O$_4$ becomes ferrimagnetic at around 860 K and remains so down to very low temperatures, all over the metal-insulator Verwey transition).\cite{verwey} 
In this case, a large magnetoelectric coupling is a-priori not to be expected, also given the different origin and ordering temperatures for the magnetic and dipolar orders.

\subsubsection{Fe$^{2+}$/Fe$^{3+}$ charge patterns in magnetite}
\label{sec:magnetite}

\begin{figure}[!h]
\resizebox{0.9\columnwidth}{!}{%
  \includegraphics{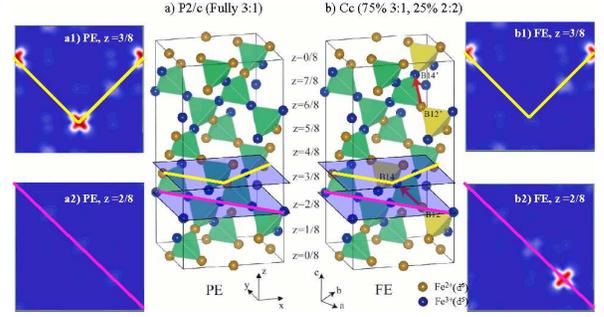}
}
\vspace*{2.5cm}       % Give the correct figure height in cm
\caption{Ionic structure of Fe octahedral sites in a) P2/c and b) Cc. 
Orange and blue balls show Fe$^{2+}$ and Fe$^{3+}$ ions, respectively. 
Fe tetrahedra of 2:2 and 3:1 CO patterns are highlighted by yellow 
and black color planes, respectively. Electric dipole moments caused 
by charge shifts are indicated by red arrows. For details, see 
Ref.\cite{yamauchi.prb}. In the four squared blue boxes, we show the charge/orbital ordering of Fe minority $t_{2g}$ states in PE and FE states at different planes for different $z$ internal coordinates in the unit cell: (upper-left, a1): PE for $z$=3/8; (lower-left, a2): PE for $z$=2/8; (upper-right, b1): FE for $z$=3/8; (lower-right, b2): FE for $z$=3/8; .}
\label{fig:magnetite}       % Give a unique label
\end{figure}

Magnetite is probably one of the  (if not ``{\em the}") most studied    magnets: it was discovered in Greece around 6$^{th}$ century before Christ and, since then, it has always attracted lots of interests. Fe$_3$O$_4$ (formally	Fe$_A^{3+}$  [Fe$^{2.5+}$Fe$^{2.5+}$]$_B$ O$_4^{2-}$)	shows an inverted cubic spinel structure with a $Fd-3m$ space group at room temperature. The inverted spinel structure shows Fe$_A$ and Fe$_B$ ion sites  coordinated to O ions, {\em i.e.} tetrahedral Fe sites are occupied by Fe$_A$ ions, whereas  octahedral Fe sites are occupied by Fe$_B$ ions. The latter form a network of corner sharing tetrahedra. Magnetic moments on Fe$_A$ sites are antiparallel to those of Fe$_B$ sites, so that ferrimagnetism is the ground state. Magnetite undergoes a first order metal-insu\-lator transition (called Verwey transition) at around
120 K\cite{verwey}, where	the	resistivity	increases	by two orders of magnitude. Correspondingly, the crystal structure changes from cubic to monoclinic. Verwey has proposed the metal-insulator transition to originate from charge ordering at Fe$_B$ sites	(Fe$_A^{3+}$  [Fe$^{2+}$Fe$^{3+}$]$_B$ O$_4^{2-}$). The pattern of the charge ordering, however, constitutes a matter of debate and it is still unknown. Anderson	has	
pointed	out\cite{Anderson} that, when putting two Fe$^{2+}$ and two Fe$^{3+}$  sites on each Fe$_B$ tetrahedron (so called ``2:2" pattern), the number of  Fe$^{2+}$-Fe$^{3+}$ ion pairs is maximized , therefore giving rise to the lowest possible energy from the Coulomb repulsion point of view. However, the Anderson criterion for low temperature magnetite is inconsistent with recent experimental results
and alternative patterns with  ``3:1'' CO arrangement (three Fe$^{2+}$ and one Fe$^{3+}$ ions in a tetrahedron, or viceversa)  or even ``mixed 75$\%$ 3:1 and 25$\%$ 2:2" have been put forward.

In our recent works, pure charge-order (CO) was carefully investigated as a potential source of inver\-sion- sym\-metry-brea\-king electronic order. Indeed, this mechanism was explored in magnetite below the Verwey metal-insulator transition. In a joint theory-experiment 
study \cite{alexe} and in following purely theoretical 
studies \cite{yamauchi.prb,tets1}, we showed that magnetite in the $Cc$ symmetry (predicted by density--func\-tional-theory (DFT) to be the ground state and suggested to experimentally occur at very low temperatures) shows a non-centrosymmetric CO of Fe$^{2+}$/Fe$^{3+}$ on octahedral Fe$_B$ sites of Fe$_3$O$_4$, with P $\sim$ 5 $\mu$C/cm$^2$. Magnetite might therefore be considered as ``one of the first multiferroics known to mankind".
Remarkably, it is a beautiful example from another point of view: since what is usually searched for in electrically-controllable spintronic devices is a {\em net magnetization}, magnetite, being a
 {\em ferri-magnet}, overcomes the limitation of a zero (and therefore uncontrollable) magnetization	that	occurs	in	many other multiferroic antiferromagnets.

As shown in Figure\ \ref{fig:magnetite}, octahedral Fe sites, arranged in the corner sharing tetrahedral network  are located in $xy$ planes with $z$=$i$/8 ($i$=0...7). The $P2/c$ paraelectric state  has ${E,C_{2b}+(0,0,1/2),I,\sigma_{2b}+(0,0,1/2)}$ symme\-tries (with   a full 3:1 tetrahedron CO arrangement) and the $Cc$ ferroelectric state has  ${E,\sigma_{2b}+(0,0,1/2)}$ symmetries (with a mixed CO pattern) in a conventional base-centered monoclinic cell so that there are two equivalent atoms (cfr B12 and B12' sites in 
Figure\ \ref{fig:magnetite}). We remark that  the mirror symmetry along with the translation vector forbids any net polarization along $b$ and finite P is allowed only along the $a$ and $c$ directions. 
The difference between the two $Cc$ ferroelectric and $P2/c$  paraelectric CO distributions (see Figure\
 \ref{fig:magnetite} a) and b)) can be understood when assuming a charge ``shift" from B12 to B14 site and in the upper part of the cell, from B12' to B14', all the other sites keeping their valence state unaltered. Each ``charge shift" creates two 2:2 CO tetrahedra, so as to form, in total, four 2:2 tetrahedra in the unit cell 
(cfr. Figure\ \ref{fig:magnetite}). The resulting CO pattern lacks inversion symmetry, therefore allowing FE polarization. 
The Berry phase approach predicts quite a large polarization, its direction lying in the $ac$-mirror plane. The DFT results are in excellent agreement with recently reported experimental values for magnetite thin films (reporting $P$ of the order of 5.5	$\mu$C/cm$^2$	in	the	$ab$	 plane	with	the	$c$	component	not measured) as well as with earlier experiments on single crystals ($P_a$ = 4.8	$\mu$C/cm$^2$	and $P_c$ = 1.5	$\mu$C/cm$^2$). We've verified that the polarization values are not largely affected by the value of the Hubbard $U$ parameter, as shown
 in Table \ref{tbl.Udependence}. What we also note is that, upon increasing U and keeping the atomic configuration fixed to that obtained for $U$=4.5 eV, the charge separation between Fe$^{2+}$ and Fe$^{3+}$ is increased, in agreement with what intuitively expected: a ``full charge disproportionation"  to occur in the limit of an infinitely large Coulomb repulsion.

\begin{table}[h] \begin{center}
\caption{Charge separation (cs, {\em i.e.} difference of $d$-charges between Fe$^{2+}$ and Fe$^{3+}$ ions in the atomic sphere with 1\AA $\:$ radius) and the corresponding FE {\bf P$_{Berry}$}  (in $\mu$C/cm$^{2}$) vs  Coulomb repulsion $U$ ($J$ is fixed to 0.89 eV).}
\label{tbl.Udependence}
\begin{tabular}{cccc}
\hline
$U$ (eV)& 4.5 & 6.0 & 8.0 \\
\hline
cs&           0.17&0.23&0.30\\
%$P$&       (-3.18, 3.18, 4.13)&(-3.13, 3.13, Nan)&(Nan, Nan, Nan)\\
%$P$&       (-4.50, 0, 4.13)&(-4.42, 0, 4.81)&( -4.33,0,5.07)\\
{\bf P$_{Berry}$} &       (-4.41, 0, 4.12)&(-4.42, 0, 4.81)&(-4.33, 0, 5.07)\\
\hline
\end{tabular}
\end{center}
\end{table}
 
We remark that, especially in CO-materials (such as magnetite), we' ve often found an antiferroelectric (AFE) phase energetically competing with a 
ferroelectric one\cite{tets1}. For example, 
in magnetite, the ground-state FE $Cc$ symmetry is
 only a few meV/unit-cell lower than AFE $P2/c$;
 also in the case of Fe$_3$O$_4$ phases in which an 
intermediate bond-and-site-centered CO occurs, 
the AFE $P2/c$ symmetry competes with the FE $P2$ symmetry\cite{tets1}.

\subsection{Orbital ordering }

In addition to  charge and spin, electronic degrees of freedom include the orbital one. Indeed, many transition metal oxides clearly show, below a critical temperature, an orbital-order (OO), often driven by the Jahn-Teller effect and accompanied by structural distortions in the octahedral or tetrahedral oxygen cages which surround transition metal ions. In principle, there seems to be no obstacle, from the symmetry point of view, to the fact that OO itself could break inversion symmetry and give rise to polarization, {\em i.e.} nothing precludes orbital-induced ferroelectricity. However, there are no established examples where this happens in a clear and simple way. For example, the Ruddlesden--Popper bilayer manganite, \\Pr(Sr$_{0.1}$ Ca$_{0.9}$)$_2$Mn$_2$O$_7$
\cite{doublelay:Tokura}, was proposed as a candidate material in this context, since  the 	rotation	of	orbital	pattern below a defined critical temperature	 was found to happen along with a ferroelectric  state. However, many degrees of freedom were active at the same time in that compound: in addition to OO and ferroelectricity,
 CO coupled with the underlying lattice distortion was also occurring,
 so that the link between OO and polarization is actually  under debate.
 A pure system in which OO by itself drives ferroelectricity is still to be found. 
 What we will discuss in the following section is, on the other hand, a compound where orbital-order occurs and, indirectly via hydrogen bonds, induces a polar state in an organic-inorganic 
hybrid\cite{mof15}. 
  
\subsubsection{Cu-based metal-organic-framework: role of orbital order and hydrogen-bond }
%%%%\label{sec:cumof}

\label{sec:mof}
Metal-organic frameworks (MOFs) are an interesting new class of materials 
made up of extended ordered networks of metal cations linked by organic 
bridges\cite{mof1}. They are hybrid organic-inorganic  materials at 
the interface between molecular chemistry and materials science. 
There is an huge interest in these materials for their potential 
technological applications such as 
gas storage, exchange or separation, catalysis, drug delivery,
optics, magnetism\cite{mof2,mof3}. Furthermore,  due to their dual nature, 
they can be engineered in almost 
infinite ways by playing with the organic/inorganic components\cite{mof4,mof5,mof6}.
A very recent family of MOFs, with a more dense topology, 
mimics the ABX$_{3}$ perovskite inorganic topology. These compounds show 
interesting magnetic, optical, electronic and dielectric properties. Last but not least, 
coexistence of ferroelectricity and magnetism, \textit{i.e.} 
multiferroicity\cite{mof7,mof8,mof9,mof10,mof11,mof12}.
\begin{figure}[!h]
\resizebox{0.5\textwidth}{!}{\includegraphics{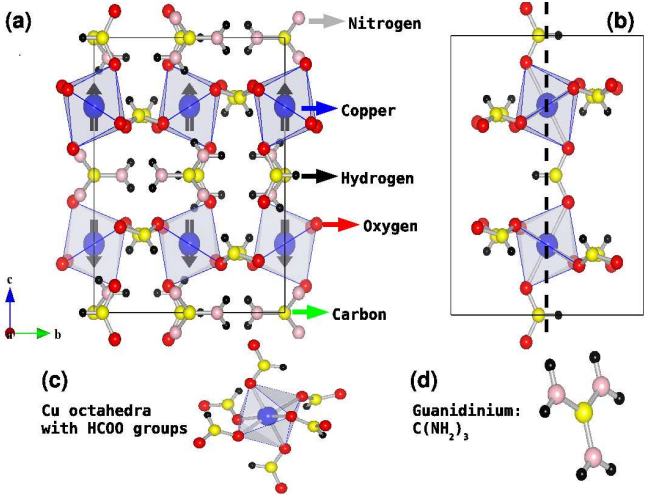}}
\caption{Crystal structure of the Cu-MOF: (a) side view, (b) a single linear chain along the $c$ polar axis, 
(c) an octahedron with the HCOO organic linkers, (d) perspective view of the Guanidinium ion.}
\label{mof_fig1}
\end{figure}

We have recently studied a Cu based MOF, namely  [C(NH$_{2}$)$_{3}$]Cu[(HCOO)$_{3}$], first 
synthesized by Ke-Li Hu \textit{et al.}\cite{mof13} In this compound, hereafter called Cu-MOF, 
A is the guanidinium ion C(NH$_{2}$)$^{+}$, 
B is the Jahn-Teller Cu$^{+2}$ ion with d$^{9}$ with t$_{2}^{6}$e$^{3}$ electronic configuration
and X is the carboxylic linker HCCO$^{-}$. At low temperature, it crystallizes in a 
polar space group  $Pna2_{1}$. Furthermore, a magnetic study revealed that it 
displays spin-canted antiferromagnetism,
with a Neel temperature of  4.6 K. 
In addition to the general
spin-canted antiferromagnetism, it 
is a magnetic system with low
dimensional character\cite{mof13}. A magnetic ordering in a polar space group 
immediately calls for a possible multiferroic behaviour, although 
no ferroelectric hysteresis loop has been measured yet. Our theoretical study 
predicts and supports a multiferroic behaviour. Furthermore it highlights 
interesting features in this appealing class of materials, such as an unusual microscopic mechanism 
for  ferroelectricity 
and the magnetoelectric effect.

In Figure\ \ref{mof_fig1}, we show: (a) side view  of the Cu-MOF; (b) a single chain of 
octahedra connected by the  HCOO  organic linkers along the polar $c$ axis; (c) Cu octahedra with HCOO groups; 
(d) perspective view of Guanidinium ion. 
The Jahn-Teller distortion of the Cu octahedra give rise to two short (s) 
and two long (l) equatorial Cu-O$_{eq}$ bonds with lengths $\sim$ 2.0\ \AA\ and 2.4\ \AA\ respectively, 
and two medium (m) apical Cu-O$_{ap}$ bonds. The cooperative Jahn-Teller distortion is characterized 
by CuO$_{6}$ octahedra elongated along the [1,1,0] and [1,1,0] in the $ab$ plane.

Our calculations show that 
the most stable magnetic configuration  
is the AFM-A type, which shows  $ab$ intra-plane ferromagnetically 
aligned spins which are inter-plane antiferromagnetically coupled.
When including   spin-orbit coupling in the calculation, 
a weak-ferromagnetic (FM) component arise due to a small canting 
of the spins. The weak-FM component M$_{a}$ is along  the $a$ axis, 
perpendicular to $c$ axis.
The presence of weak-ferromagnetism  is in agreement with 
experimental observation\cite{mof13}.
The perovskite Cu-MOF is very similar to KCuF$_{3}$
which is 
considered as a prototypical system for a cooperative
Jahn-Teller effect, orbital ordering, 
and a quasi-one-dimensional antiferromagnetic
Heisenberg chain. In fact, the Cu-MOF shows a 
particular type of orbital order,
in which a single hole alternately occupies 
3d$_{x^{2}-z^{2}}$ and 3d$_{y^{2}-z^{2}}$
orbital states of
the Cu$^{+2}$ ions 
(3d$^{9}$ electronic configuration)\cite{mof14}.
The cooperative Jahn-Teller
distortion is characterized by Cu(HCOO)$_{6}$ 
octahedra alternatively elongated along the perpendicular 
[1,1,0] or [$\bar{1}$,1,0]
directions in the $ab$-plane, i.e. giving rise to an 
antiferrodistortive pattern. It is important to note that 
the anti-ferro-distortive (AFD) modes are usually non-polar
 distortions in standard inorganic perovskite like compounds,
 and, as such, they 
should not give rise to ferroelectric polarization. Despite this, our calculations 
show that the AFD distortions in Cu-MOF are strictly correlated to the presence of the polarization. 

Our study highlights very interesting properties 
of this compound:\\
{\em i)} it is ferroelectric with an estimated polarization $P$ of 0.37
 $\mu C/cm^2$, with polar axis along $c$;\\
{\em ii)} the microscopic mechanism is very intriguing: 
we found that non-polar AFD distortions are  
intimately related to ferroelectric polarization suggesting that 
they may be the ``source" of ferroelectricity;\\
{\em iii)} inspection into the microscopic mechanism of {\em ii)} shows 
that AFD distortions acting on the BX$_{3}$ framework 
are coupled to A-group ions through intervening 
hydrogen-bonds between the Oxygens of the Cu(HCOO)$_{6}$
and the H atoms of the A-group.
While the AFD distortions alone would preserve centrosymmetry, 
the O$\cdots$H bonds induce asymmetric distortions into 
the A-group, ultimately responsible of the presence of 
  dipoles mainly localized at the A-group  which, in turn, give rise to a finite polarization;\\
{\em iv)} the weak-ferromagnetic component is strictly correlated with the 
ferroelectric polarization: when P is equal to zero,
 the weak-FM components goes
to zero; at the $+P$ state, it is $+M_{a}$ and at the $-P$ state it is 
$-M_{a}$. Therefore, 
 our calculations predicts that the Cu-MOF should be 
a magnetoelectric multiferroic: 
it should be possible to control the magnetization 
by an external electric field. 
In particular, an electric field along the $c$ axis, which 
would switch the spontaneous polarization, would at the same time,  
switch the sign of $M_{a}$. Although  both $P$ and $M$ are 
small in magnitude, this opens new avenues in the multiferroic research 
in such a novel and exciting class of materials. 
It goes without saying that there could be large room 
for engineering these compounds for enhancing 
these magnetoelectric effects due to the organic-inorganic duality characteristic 
of MOFs. More discussions about the Cu-MOF can be 
found in Ref.\cite{mof15}

In conclusion, MOFs  are materials at the  border-line between
 chemistry and solid state physics and MF-MOFs
represents a ``dual-bridge" between the two fields, exploiting 
knowledges from the inorganic as well as organic material science. 
We expect that this dual-bridge will be the source of new and 
interesting physical properties
\cite{mof16,mof17},
 which ab-initio studies can easily unveil\cite{mof15}.
Incidentally, we note that ab-initio characterization of MF-MOFs are almost 
totally lacking in the current literature, 
and our recent study\cite{mof15} is certainly encouraging
 in terms of interesting results.

\subsection{Charge-spin dimers in donor-acceptor TTF-CA}
\label{sec:ttfca}

%\label{sec:TTFCA}
In comparison with inorganic materials, organic compounds
have been synthesized in large number but ferroelectric properties
have been found only rarely in that class of materials\cite{ttfca1}.  A breakthrough in organic ferroelectricity was recently achieved by  the discovery of very large room-temperature ferroelectric polarization  in  the croconic acid, a well-known low-molecular-weight 
organic compound\cite{nature}. It is believed  that it may be fruitful to search among known - but poorly
characterized - organic compounds for organic ferroelectrics with enhanced polar properties suitable for device 
applications\cite{ttfca3}.

Coexistence of ferroelectric and magnetic order in organic materials is an even rarer property.  Recently, we have predicted by ab-initio calculations that multiferroicity 
may be found in TTF-CA molecular crystal\cite{ttfca}, a multi-component molecular system
which produces a typical displacive-type ferroelectricity by displacing
oppositely charged species. TTF-CA is a  charge-transfer (CT) complex
composed of electron donor (D) and acceptor (A) molecules, such as tetrathiafulvalene (TTF)
and tetrachloro-$p$benzoquinone (CA)\cite{ttfca5,ttfca6,ttfca7,ttfca8,ttfca9,ttfca10,ttfca11,ttfca12,ttfca13,ttfca14,ttfca15}. 
This compound is particularly interesting because 
it shows a neutral-ionic (NI) phase transition, \textit{i.e.} a transition between a van der Waals molecular assembly 
to an ionic solid\cite{ttfca8,ttfca9}. The ionized molecules form DA dimers, D$^{+\rho}$-A$^{-\rho}$, where $\rho$ is a degree of charge transfer, with a lowering of the crystal structure from $Pn$ to a polar $P2_{1}/n$ space group, where the originally non polar D$\cdots$A$\cdots$D$\cdots$A sequence with regular intermolecular separation are symmetry broken to a polar chain formed by the DA dimers characterized by the formation of pairs of short and long bonds along the stacking axis $a$. In essence, above the NI transition temperature T$_{NI}$ of $\sim$ 
84 K\cite{ttfca9} the system is in a neutral and
 paraelectric state
 with D$^{+\rho}$$\cdots$A$^{-\rho}$$\cdots$D$^{+\rho}$$\cdots$A$^{-\rho}$ with $\rho$=0.2-0.3. Below T$_{NI}$ the system becomes ferroelectric  with a stacking D$^{+\rho}$$\cdots$A$^{-\rho}$$\cdots\cdots$D$^{+\rho}$$\cdots$A$^{-\rho}$ with $\rho$$\sim$ 0.6, with an ionic and ferroelectric state characterized by a Pierls-like dimerization.

In this framework, we have performed ab-initio calculations
 by using the recently introduced screened hybrid functional
 Heyd-Scuseria-Ernzerhof (HSE)\cite{ttfca}. The use of the hybrid functional has been particularly important here for several reasons. 
First, 
it is important to improve the description of the HOMO-LUMO gap which governs the degree of charge transfer between molecular units; second, we found it impossible to stabilize a magnetic 
state by using the local or semilocal approximation to the exchange-correlation functional (LDA or GGA); third, the commonly used DFT+$U$ method for improving the electronic structure of ''strongly`` correlated system can not be directly applied here: the reason is that
in molecules, the localized orbitals are multicenter rather than single-center, since the basis set correspond to ortho-normal molecular orbitals instead of orthonormal atomic
orbitals. The main results of our study can be summarized as follows: 
the HSE ground state of the TTF-CA crystal shows an antiferromagnetic (AFM) ordering, more stable than a non-magnetic one by $\sim$ 80 meV per unit cell. Note that starting with an initial ferromagnetic (FM) configuration, the solution converges again to an AFM one. This demonstrates the robustness of our AFM solution. The single TTF and CA units become spin-polarized with a net polarization of $\sim$ 0.40 $\mu_{B}$ per molecule. The HSE energy gap of spin-polarized dimerized state is 0.5 eV.  A ferroelectric state with a coexisting magnetic ordering as ground state characterizes the TTF-CA as the 
first multiferroic organic crystals, predicted by ab-initio calculations. The calculated ferroelectric polarization in the AFM state is  3.5 $\mu C/cm^{2}$ while in the NM state is 
8.0 $\mu C/cm^{2}$. The sensitivity of the polarization to  the magnetic state is mainly due 
due to the large increase of the electronic component of polarization upon changing from the 
NM to AFM state. The ionic one, on the other hand, does not depend much on the magnetic state, and it is always opposite to the electronic one. Further details can be found in 
the original article\cite{ttfca}. Finally, 
the multiferroicity in TTF-CA has been independently
 confirmed by theoretical calculations\cite{ttfca16}. 
A recent experimental result show that  one-dimensional quantum magnets, such as organic charge-transfer complexes, could be promising candidates in the development of magnetically controllable ferroelectric 
materials\cite{ttfca17,ttfca18,ttfca19,ttfca20}.

\subsection{Coupled distortions in NaLaMnWO$_6$}
%%%\label{sec:NLCWO}

\label{sec:NaLaMnWO6}

As well discussed so far, perovskite  (with formula ABX$_{3}$)  is one of the crystalline structures which is most commonly
occurring and most important in all of materials science.  Because of the great flexibility inherent in the perovskite structure, mainly due to the corner sharing octahedra,
there are many different types of distortions which  occur starting from the ideal cubic structure.
These include tilting of the octahedra, displacements of the cations out
of the centers of their coordination polyhedra
and distortions of the octahedra driven by electronic factors ({\em i.e.} Jahn-Teller distortions). 
Many of the physical properties of perovskites depend crucially on the details of these distortions, particularly the electronic, magnetic and dielectric properties which are so important for many of the applications of perovskite materials.

The new class of double perovskites AA'BB'O$_{6}$ introduces yet another degree
of freedom, namely the possibility of cation ordering on both A and B sites.
Obviously, this  greatly increases the possibility of functional design in this class
of compounds. More than  20 new examples of this structure type have been discovered so far. These materials are found to have highly complex microstructures and show potential for
 multiferroic behavior\cite{qq1,qq2,qq3,qq4,qq5,qq6}.

We have recently presented a theoretical study of the structural and ferroelectric properties
of the new double-perovkite NaLaMnWO$_{6}$\cite{qq7,qq8,qq9}, which belongs to this class of materials, 
by combining group-theoretical analysis and first-principles calculations to explore the origin of the polar state in this compound.  NaLaMnWO$_{6}$  orders magnetically at low temperature
in a polar space group. However, the ferroelectricity has neither been calculated nor measured
yet.  We found  that ferroelectricity originates
not from a usual type of lattice distortion involving 
small off-centerings of ions, as usually occurring in the prototypical ferroelectric BaTiO$_3$, but from the combination of two oxygen rotational
distortions. 

This idea of rotation driven ferroelectricity is a very exciting recent development in the field of ferroelectrics as well as in  
the related field of multiferroics.   This represents an interesting new route 
to produce new multiferroic and magnetoelectric materials, relying on the idea 
of starting with non-polar materials and then induce multiple non-polar instabilities;
under appropriate circumstances, this can induce a ferroelectric polarization, 
as first predicted in Ref.\cite{qq10} based
on general group theory arguments and analyzed in the
SrBi$_{2}$Nb$_{2}$O$_{9}$ compound by means of a symmetry analysis
combined with density-functional theory calculations by
Perez-Mato \textit{et al.}\cite{manuel}. In that case,  ferroelectricity was found to
arise from the interplay of several degrees of freedom, not
all of them associated with unstable or nearly-unstable
modes. In particular, a coupling between polarization
and two octahedral-rotation modes was invoked to explain 
the behavior\cite{manuel}. Bousquet {\em et al.} have demonstrated
that ferroelectricity is produced by local rotational modes
in a SrTiO$_{3}$/PbTiO$_{3}$ superlattice\cite{qq11}. Benedek and Fennie 
proposed that the
combination of two lattice rotations, neither of which produces
ferroelectric properties individually, can induce a ME coupling, weak
ferromagnetism, and ferroelectricity\cite{Ref33}.
Indeed, we now know
that rotations of the oxygen octahedra, in
combination\cite{Ref33,Ref34,Ref35} and even
individually\cite{ederer06,Jorge1}, can produce ferroelectricity,
modify the magnetic order, and favor magnetoelectricity.

\begin{figure}[!h]
\resizebox{0.5\textwidth}{!}{\includegraphics{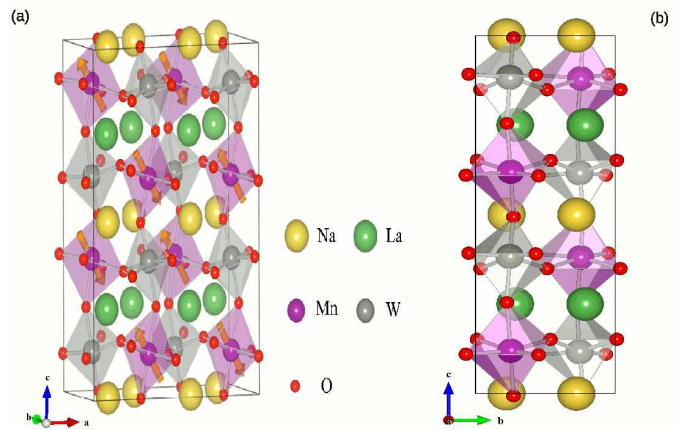}}
\caption{Crystal structure of the NaLaMnWO$_{6}$ compound: 
(a) perspective view and  (b) side view. The spins are shown as arrows only 
in (a).}
\label{NaLa}
\end{figure}

Our study on NaLaMnWO$_{6}$
 represents another step forward along this new emerging direction.
 In Figure \ref{NaLa} (a) we show a perspective (a) and side (b) 
view of the magnetic unit cell of the compound.  
In ref.\cite{nlmwo} we showed that this compound is a potentially very
 interesting multiferroic compounds for several reasons:
{\em i)} the estimated polarization is very large, about 16 $\mu C$/cm$^{2}$;
{\em ii)} an intriguing mechanism is at the basis of the ferroelecticity:
two primary non-polar distortions such as tilting and rotation of octahedra - typical
of  perovskite systems - in combination with cation ordering induce the  breaking
of inversion symmetry and allows for a ferroelectric polarization.
By comparing the low symmetry structure with a
parent phase of P4/nmm symmetry, 
two distortion modes are found dominant. They correspond to
MnO$_{6}$ and WO$_{6}$ octahedron tilt modes, 
often found in many simple perovskites. While in the latter
these common tilting instabilities yield non-polar phases, 
in NaLaMnWO$_{6}$ the additional presence of the A-A$^{'}$ 
cation ordering is sufficient to make these rigid unit modes 
as a source of improper
ferroelectricity. Through a trilinear coupling with
 the two unstable tilting modes, a  polar
distortion is induced: 
  a negligible polar instability does exist, 
but the additional A cation layer ordering makes
 ferroelectrically active some tilting modes
of the octahedra that in simple perovskites and in 
B-ordered double perovskites only give rise to non-polar
behaviour.  Despite its secondary
character, this polarization is 
coupled with the dominant tilting modes and 
its switching is bound
to produce the switching of one of two tilts, 
enhancing in this way a possible interaction with the
magnetic ordering. Through a trilinear coupling with the two unstable 
tilting modes, a significant polarization is induced.
We hope that this study will stimulate further investi-
gation of cation ordering as a tool to convert ubiquitous
well-known steric non-polar instabilities into mechanisms
for producing improper ferroelectrics, as well as new multiferroics.
Further details can be found in Ref.\cite{nlmwo}.

\section{Conclusions and perspectives}
\label{sec:concl}

It is  evident that multiferroics   currently represent   a vivid and promising field of materials science, the enthusiasm being mainly driven by {\em i}) the impression  that multiferroics are much more common than what originally thought and {\em ii}) the rich and still largely unexplored variety of mechanisms leading to the coexistence of multiple orders. Density functional theory is able to identify the microscopic mechanisms, their strengths as well as their limitations, their chemical and physical origin, so it seems a  particularly suited technique for the analysis of this complex class of materials.

As shown in this paper, our recent activity was devoted to those materials where ferroelectricity is induced by  peculiar charge, spin or orbital orders which lack inversion symmetry, {\em i.e.} to the so called ``improper electronic ferroelectrics". We'll try in this conclusive paragraph  to summarize 
the main findings of our recent activity as well as to give guidelines towards an efficient materials-design for optimized multiferroics. 

\begin{itemize}

\item When dealing with spin-driven ferroelectricity, 
two mechanisms have been mainly explored so far: the first one is based on Heisenberg exchange coupling, whereas the second one is based on relativistic Dzyaloshinskii-Moriya interaction. An important issue regards the efficiency of these mechanisms, or equivalently, the magnitude of the  polarization induced by the two mechanisms. According to our estimates based on density functional theory, the 
polarization caused by Heisenberg exchange (at play in collinear spin configurations) is much larger than that driven by Dzyaloshinskii-Moriya exchange (at play in non-collinear spin configurations). This was quantitatively shown, for example, in 
nickelates \cite{nickel}, when we compared 
first--principles estimates obtained for polarization in both  collinear or spiral spin arrangements, as
 experimentally proposed\cite{nickexp}, the two values differing by approximately two orders of magnitude.  This is consistent with what expected, based on the argument that relativistic effects are not very sizeable in 3$d$ transition metal oxides and that the symmetric exchange coupling is much larger than the corresponding antisymmetric component.

\item Many systems were studied, all of them showing Heisen\-berg-driven 
polarization\cite{prlslv,tets,nickel}.
The magnitude of the latter, as expected, is strongly dependent on the involved transition metal and spin-state. 
In a rather qualitative, general and naive way, one can invoke the size of spin moments and exchange-interactions to rationalize the different behaviour in different oxides.
In particular, we have quantitatively estimated the biggest effect to occur in rare-earth manganites involving Mn$^{3+}$ ($d^4$), with polarization of the order of $\mu C/cm^2$. This big value is reasonably due to the large size of the spin moment ($\sim$ 4 $\mu_B$) and  to rather strong interactions between Mn $e_g$ and oxygen $p$ states which are  able to well mediate the Mn-Mn exchange interaction in ortho-manganites. As a result, the (large) generic Heisenberg term, $J_{i,j} \: S_i \cdot S_j$,  can  induce appreciable changes in the energy and related sizeable (local) distortions, depending on whether $S_i$ and $S_j$ spins are parallel or antiparallel.  A large polarization is therefore expected when all the (local) distortions are summed up over the magnetically-ordered unit cell, given the overall polar distortion pattern and non-centrosymmetric spin configuration. For a similar reason (but producing an opposite result), the V-V dimerization is able to induce a much smaller polarization (order of few tenths of a $\mu C/cm^2$) in spinel Cd-based vanadate 
(cfr Sec.\ref{sec:cvo}): spin moments are smaller in size and the states at play are $t_{2g}$ ({\em i.e.} much less prone than $e_g$ to interactions with oxygen, due to their main non-bonding character, with likely smaller exchange constants and related smaller distortions) . Also in the case of spin-driven ferroelectricity for $f-d$ systems, of which the prototypical DyFeO$_3$  
case was discussed in Section\ \ref{sec:dfo},  one can expect a smaller exchange interaction between {\em localized} 4$f$ states and ``{\em semi--localized}" 3$d$ states, compared to the exchange interaction between 3$d$ states. Since it is the $f-d$ coupling that induces the polar configuration in DyFeO$_3$ and  the spins have to order on {\em both} Dy and Fe sublattices, we expect a smaller polarization than in, say, orthomanganites, as indeed predicted  by our {\em ab--initio} simulations. 
Incidentally, we note that spin-driven ferroelectricity based on 4$f$ states is likely to develop only at  temperatures as low as $\leq$ 10 K, where the rare-earth ions order, as is the case of DyFeO$_3$  

\item How to increase the ordering temperatures for magne\-tical\-ly-induced ferroelectrics (commonly of the order of few tens of Kelvins) represents in general one of the toughest challenges towards finding a so-called ``killer-app" in the field of multiferroics for them to  become really technologically appealing. Although progresses were made in recent years  (for example, by focusing on combined charge- and spin-ordered materials, 
such as nickelates\cite{nickel} and hole-doped
 manganites\cite{GGhalfdoped}, showing ordering temperatures of 150-200 K),  room temperature operation is still a dream. One of the limitations might be constituted by the fact that electronic magnetic ferroelectrics are generally {\em frustrated} materials (showing either spin or charge or orbital frustration), whose ordering temperatures are - due to competing interactions - intrinsically small. The possible way out could be to deal with large exchange-coupling constants, which is what happens   
in nickelates and in CuO tenorite\cite{cuo,prljose}. The latter material, in particular, offers an example of spin-spiral-based multiferroic with a large ordering temperature (230 K). 

\item When dealing with CO-induced ferroelectricity, polar CO (as identified in Fe$^{2+}$/Fe$^{3+}$ occurring in
 magnetite\cite{yamauchi.prb}, Mn$^{3}$+/Mn$^{4+}$ in half-doped 
manganites\cite{GGhalfdoped}, Ni$^{2+}$/Ni$^{4+}$ in rare-earth 
nickelates\cite{nickel}) can lead to potentially large ferroelectric polarization (of the order of few $\mu$C/$cm^2$). In this respect, CO-driven ferroelectricity is an efficient mechanism and more work towards a better understanding should be definitely carried out. There is  an important distinction to be made as for the microscopic origin of ferroelectricity. In nickelates and manganites, multiferroicity shows up when {\em both} spin and charge orders occur: it is actually their combination that drives ferroelectricity. In this case, magnetism has a relevant role in the development of the dipolar order (actually, magnetism and ferroelectricity share the same physical origin) and a large magnetoelectric (ME) coupling is therefore expected. In magnetite, on the other hand, where CO alone drives ferroelectricity, magnetism seems not to be involved in the ferroelectric transition (Fe$_3$O$_4$ becomes ferrimagnetic at around 860 K and remains so down to very low temperatures, all over the metal-insulator Verwey transition). In this case, a large magnetoelectric coupling is a-priori not to be expected, also given the different origin and ordering temperatures for the magnetic and dipolar orders. A possible reason for the magnetoelectric coupling to occur (as reported in a very 
early study back into 1994 \cite{miya}, involves the relativistic spin-orbit coupling in
 the peculiar $Cc$ symmetry\cite{maememag}

\item As a general rule-of-thumb for a larger CO-induced polarization, we remark that, in the case of non-centro\-symmetric CO, the charge disproportionation (CD) should be maximized. In fact,  according to a picture based on point-charge dipoles, valid mostly for systems where the bond is largely ionic, this would guarantee a larger polarization. To our  experience, we focused 
 on oxides (magnetite, Fe$_3$O$_4$)\cite{alexe,yamauchi.prb,tets1} and fluorides 
(K$_{0.6}$Fe$^{2+}_{0.6}$Fe$^{3+}_{0.4}$F$_3$)\cite{ttb}, both with Fe$^{2+}$/Fe$^{3+}$ charge disproportionation: the comparison between magnetite and fluorides showed that a larger CD occurs when iron is bonded to Fluorine rather than to Oxygen. Therefore, choosing a more ionic compound seems promising to achieve a large CD and related higher polarization.

\item On the theory-side, as from the methodological point of view, we remark that the treatment of correlation effects is often important to get a quantitatively reliable description of multiferroics. In addition to the common DFT+U approach, that we have
 used in a variety of studies\cite{nickel,yamauchi.prb,GGhalfdoped}, we have carried out careful simulations of prototypical multiferroics, such as the ``proper" BiFeO$_3$ and the ``improper" AFM-E HoMnO$_3$, using a state-of-the-art hybrid exchange-correlation functional, obtained by mixing the non-local Fock-exchange with a ``standard" parametrized exchange-functional,
 with encouraging results\cite{compu11}.
In particular, \\hybrid-functionals with the ``ideal" mixing between 3/4 local and 1/4 non--local exchange--corre\-laton potential, appear to give an accurate description of structural, electronic, ferroelectric, magnetic 
properties for most of the studied materials.  
This same technique was later applied to the TTF-CA organic\cite{ttfca} and DyFeO$_3$ 
multiferroics\cite{njp}. A correct description of the vibrational as well as spin-phonon coupling within hybrid functionals was also recently shown for many 
well-characterized oxides\cite{spinphonon}.
 
\item We have abundantly shown that ferroelectricity driven by electronic degrees of freedom can occur (and has actually been experimentally observed) in many systems where spin-order and charge-orders drive the rising of polarization. On the other hand, ferroelectricty induced by orbital order has remained for long elusive. 
In Sec.\ref{sec:mof} and in Ref.\cite{mof15}, we focused on a class of materials called metal-organic frameworks, i.e. corresponding organic-inorganic hybrids of perovskite crystals. 
This architecture is much more flexible (due to organic groups instead of single oxygen anions) and chemically more rich than usual inorganic perovskites (in terms, for example, of organic polar or non-polar groups 
which occupy the empty site corresponding to the A-site cation). This chemical richness suggests that new mechanisms might arise in this class of materials.
In particular, we considered a MOF based on Cu$^{2+}$ 
ions at the center of octahedral cages of COOH- groups 
and with guanidium molecules occupying A sites. 
A delicate interplay between Jahn-Teller distortions 
around Cu$^{2+}$ (in turn related to the antiferrodistortive 
orbital-ordering) and hydrogen bonding with guanidinium groups 
induces a small ferroelectric polarization. Moreover, 
following a trilinear coupling between magnetization,
 antiferromagnetism and polarization allowed by 
symmetry in the ferroelectric crystal, we reported a
 linear proportionality between weak-magnetization 
(induced by Dzyaloshinskii-Moriya coupling) and 
polarization, pointing to the long-sought electrical 
control of magnetization. 

\item Most of the  multiferroics discovered so far are antiferromagnets (due to usual strong superexchange in oxides which often favors antiparallel spins), so that their technological appeal is poor. In order to overcome this limitation, we mention two possible solutions: {\em i}) choose a compound where {\em ferrimagnetism} (quite a common spin configuration, occurring in spinels, such as magnetite, or in double perovskites, etc) is the magnetic ground state, so as to show a {\em net} magnetization that can be well controlled via a magnetic field. {\em ii}) consider (multi\-ferroic)-antiferromagnets exchange-linked to ferromagnets, so as to build an artificial heterostructure where both electric and magnetic degrees of freedom
 are simultaneously active\cite{rond1,rond2,rond3}. There, the phenomenon of exchange-bias can be used, for example, to control the magnetization  of a FM overlayer by means of an electric field which primarily modifies the ferroelectric as well as the antiferromagnetic properties of a multiferroic layer, which the FM overlayer is, in turn, exchange-coupled to (proposals for applications in this direction
 already came\cite{ramesh}).

\item The TTF-CA donor-acceptor organic crystal was probably one of the first examples of organic ferroelectrics treated from 
first-principles\cite{ttfca} (incidentally, we remark that a breakthrough in the field was
 later achieved in 2010\cite{nature}, when the croconic acid in crystalline form was discovered to be ferroelectric with large polarization persisting at least up to 400 K, showing an excellent qualitative and quantitative agreement between ab-initio theory and experiments). When looking at TTF-CA, the combination of charge transfer and structural dimerization results in an opposite behavior for the electronic and ionic contribution to polarization, which was first predicted from first--principles and later confirmed by experiments. The field of organic crystals might be richer of ferroelectrics than what originally thought, and efforts should be devoted in the near future to investigating polarization in many charge-transfer salts, charge-ordered systems and other stron\-gly-correlated organics. Proposals towards this direction already appeared in the literature, for example pointing to the quasi-two dimensional organic salt 
$\alpha$-(BEDT-TTF)$_2$I$_3$\cite{bedt}.

\item Among the many different routes to new multiferroics, particular interest has been raised by the idea of a trilinear coupling among polarization and different octahedral distortions. The idea has been originally proposed  for
ferroelectric Aurivillius
 compounds\cite{manuel} and recently rediscovered in the 
context of layered manganites, such
 as Ca$_3$Mn$_2$O$_7$\cite{craig}. 
In the present work, we've suggested yet another possibility of 
trilinear coupling in double-perovskites with formula AA'BB'O$_{6}$, 
based again on functional octahedral distortions as well as
 cation ordering and resulting in a large polarization\cite{nlmwo}.

\end{itemize}

The field has to face several challenges in the coming years. 
While electronic ferroelectrics show at least two characteristics which are definitely appealing for technological applications ({\em i.e.} expected giant magnetoelectric coupling and expected ultrafast switching), the real bottleneck is represented by the fact that working temperatures of the most studied ``improper" multiferroics are too small for applications and efforts should be devoted to 
the increase of  operating temperature range. 
%is one of the most difficult objectives to achieve in the field of 
%electronic ferroelectrics , where   frustration is a 
%main characteristic: it is, indeed, the complex charge or spin order that induces a polar charge and, since frustration is generally associated to low temperature,  the latter is a sort of ``intrinsic" limitation in this context.

An alternative way - much closer  to applications - might be represented 
by interfaces between prototypical ferromagnets and prototypical ferroelectrics - both with high operating temperatures - combined in artificial heterostructures where
the magnetoelectric coupling could be engineered and optimized. This vivid field has not been treated in the present review, which was mostly devoted to ``bulk" multiferroics; however, it should be kept in mind that many progresses were made in recent years on this kind of artificial systems 
(such as Fe/BaTiO$_3$\cite{barthefebto}), aimed in the end at implementing multiferroic memories.

Possibilities to overcome the present limitations in bulk electronic ferroelectrics might involve either the discovery of
 new physical mechanisms  in known materials or the optimization of  known mechanisms (such as spin or charge-order induced ferroelectricity found for transition-metal oxides) in ``new"  materials (such as organic or organic-inorganic hybrids, novel complex oxides). In general, a better understanding of spin--phonon coupling in transition metal oxides through first--principles approaches might certainly help in designing materials with large polarization and strong magnetism (possibly
 ferromagnetism)\cite{spinphonon}. In this respect, some works have  recently  investigated the possibility of ferroelectric and ferromagnetic instabilities in transition-metal-oxides with $d^3$ cations under strain or volume expansion 
({\em i.e.} Ca--based 
manganites\cite{rabecamno3,paolo,nicolacamno3,spinphonon},
 or La-based cromites\cite{claude}). 
 
 In summary, the physics of electronic ferroelectrics is rich and complex, so that surprises in both mechanisms and materials are to be expected in  the coming years . In this respect, we remark that efficient and reliable 
 modeling approaches can greatly contribute to the field, by proposing new materials, new mechanisms and their quantitative estimates. We therefore hope that the present manuscript will contribute to stimulate further scientific interests from the experimental point of view towards this peculiar class of materials .
 
 \section*{Acknowledgments}
The  research  has received funding by the European Community's Seventh Framework Programme FP7/2007-2013 through the European Research Council under grant agreement No. 203523-BISMUTH.
We acknowledge Kunihiko Yamauchi, Paolo Barone, Tetsuya Fukushima, Gianluca Giovannetti, Daniel Khomskii,  Jeroen van den Brink, Sanjeev Kumar,  and Elbio Dagotto  for useful discussions. A.S. would like to express 
his special thanks to Prof. J.M. Perez-Mato for his 
fruitful, deep and exciting discussions about symmetry 
aspects of the problems. A.S. thanks also Prof. H. Stokes for kind assistance for advanced applications of ISODISTORT tools. Support from the CASPUR Supercomputing Center in Rome is gratefully acknowledged.

%%%%%     \begin{table}
%%%%%    \caption{Please write your table caption here}
%%%%%    \label{tab:1}       % Give a unique label
%%%%%    \begin{tabular}{lll}
%%%%%    \hline\noalign{\smallskip}
%%%%%    first & second & third  \\
%%%%%    \noalign{\smallskip}\hline\noalign{\smallskip}
%%%%%    number & number & number \\
%%%%%    number & number & number \\
%%%%%    \noalign{\smallskip}\hline
%%%%%    \end{tabular}
% Or use
%%%%%    \vspace*{5cm}  % with the correct table height
%%%%%    \end{table}
%
% BibTeX users please use
% \bibliographystyle{}
% \bibliography{}

\end{document}